\documentclass[twocolumn]{aastex631}

\usepackage{silence}
\usepackage{enumitem}
\usepackage{amsmath}
\usepackage{mathrsfs}
\usepackage{comment}
\usepackage{fontawesome5}
\usepackage[mathlines]{lineno}
\usepackage{bbm}
\usepackage{tikz}
\usepackage{etoolbox}

\newcommand{\github}[1]{%
   \href{#1}{\faGithubSquare}%
}

\shortauthors{Sasseville et al.}
\graphicspath{{./}{figures/}}

\begin{document}
\title{Probabilistic Interpolation of Sagittarius A*'s Multi-Wavelength Light Curves Using Diffusion Models}

\author[0000-0001-8845-2025]{Gabriel Sasseville}
\affiliation{Department of Physics, University of Montreal, Montreal, QC H2V 0B3, Canada}
\affiliation{Ciela - Montreal Institute for Astrophysical Data Analysis and Machine Learning, Montreal, QC H2V 0B3, Canada}
\affiliation{Centre de recherche en astrophysique du Québec (CRAQ), Canada}

\author[0000-0001-7271-7340]{Julie Hlavacek-Larrondo}
\affiliation{Department of Physics, University of Montreal, Montreal, QC H2V 0B3, Canada}
\affiliation{Ciela - Montreal Institute for Astrophysical Data Analysis and Machine Learning, Montreal, QC H2V 0B3, Canada}
\affiliation{Centre de recherche en astrophysique du Québec (CRAQ), Canada}

\author[0000-0001-6803-2138]{Daryl Haggard}
\affiliation{Ciela - Montreal Institute for Astrophysical Data Analysis and Machine Learning, Montreal, QC H2V 0B3, Canada}
\affiliation{Centre de recherche en astrophysique du Québec (CRAQ), Canada}
\affiliation{Department of Physics, McGill University, 3600 rue University, Montréal, QC H3A2T8, Canada}
\affiliation{Trottier Space Institute at McGill, 3550 rue University, Montréal, QC H3A2A7, Canada}

\author[0000-0001-8806-7936]{Alexandre Adam}
\affiliation{Department of Physics, University of Montreal, Montreal, QC H2V 0B3, Canada}
\affiliation{Ciela - Montreal Institute for Astrophysical Data Analysis and Machine Learning, Montreal, QC H2V 0B3, Canada}
\affiliation{Mila - Quebec Artificial Intelligence Institute, Montreal, Canada}

\author[0000-0002-2603-6031]{Hadrien Paugnat}
\affiliation{Department of Physics and Astronomy, UCLA, Los Angeles, CA 90095-1547, USA}

\author[0000-0003-2618-797X]{Gunther Witzel}
\affiliation{Max Planck Institute for Radio Astronomy, Bonn 53121, Germany}



\begin{abstract}

Understanding the variability of Sagittarius A* (Sgr A*) requires coordinated, multi-wavelength observations that span the electromagnetic spectrum. In this work, we focus on data from four key observatories: Chandra in the X-ray (2–8 keV), GRAVITY on the Very Large Telescope in the near-infrared (2.2 $\mu m$), Spitzer in the infrared (4.5 $\mu m$), and ALMA in the submillimeter (340 GHz). These multi-band observations are essential for probing the physics of accretion and emission near the black hole’s event horizon, yet they suffer from irregular sampling, band-dependent noise, and substantial data gaps. These limitations complicate efforts to robustly identify flares and measure cross-band time lags, key diagnostics of the physical processes driving variability. To address this challenge, we introduce a diffusion-based generative model, for interpolating sparse, multivariate astrophysical time series. This represents the first application of score-based diffusion models to astronomical time series. We also present the first transformer-based model for light curve reconstruction that includes calibrated uncertainty estimates. The models are trained on simulated light curves constructed to match the statistical and observational characteristics of real Sgr A* data. These simulations capture correlated multi-band variability, realistic observation cadences, and wavelength-specific noise. We compare our models against a multi-output Gaussian Process. The diffusion model achieves superior accuracy and competitive calibration across both simulated and real datasets, demonstrating the promise of diffusion models for high-fidelity, uncertainty-aware reconstruction of multi-wavelength variability in Sgr A*.

\end{abstract}



\section{Introduction} \label{sec:intro}

In recent years, the field of time-domain astrophysics has undergone rapid growth, driven by increasingly sophisticated observational facilities and methods (e.g., \citealt{bellm_zwicky_2019, ivezic_lsst_2019}). Time-domain astrophysics investigates transient objects and phenomena that change over time on timescales ranging from milliseconds to years (e.g., \citealt{vries_structure_2005, zhang_physics_2023}). These studies are critical for understanding processes in a wide range of astrophysical systems, from fast radio bursts to accreting black holes (BHs; e.g., \citealt{zhang_physics_2023, fan_quasars_2023}). By analyzing how these sources vary over time, we can gain insights into the underlying physical mechanisms driving their behavior such as accretion and ejection processes, jet formation and magnetic field dynamics (e.g., \citealt{boyce_multiwavelength_2022}).

\begin{figure*}[ht]
\plotone{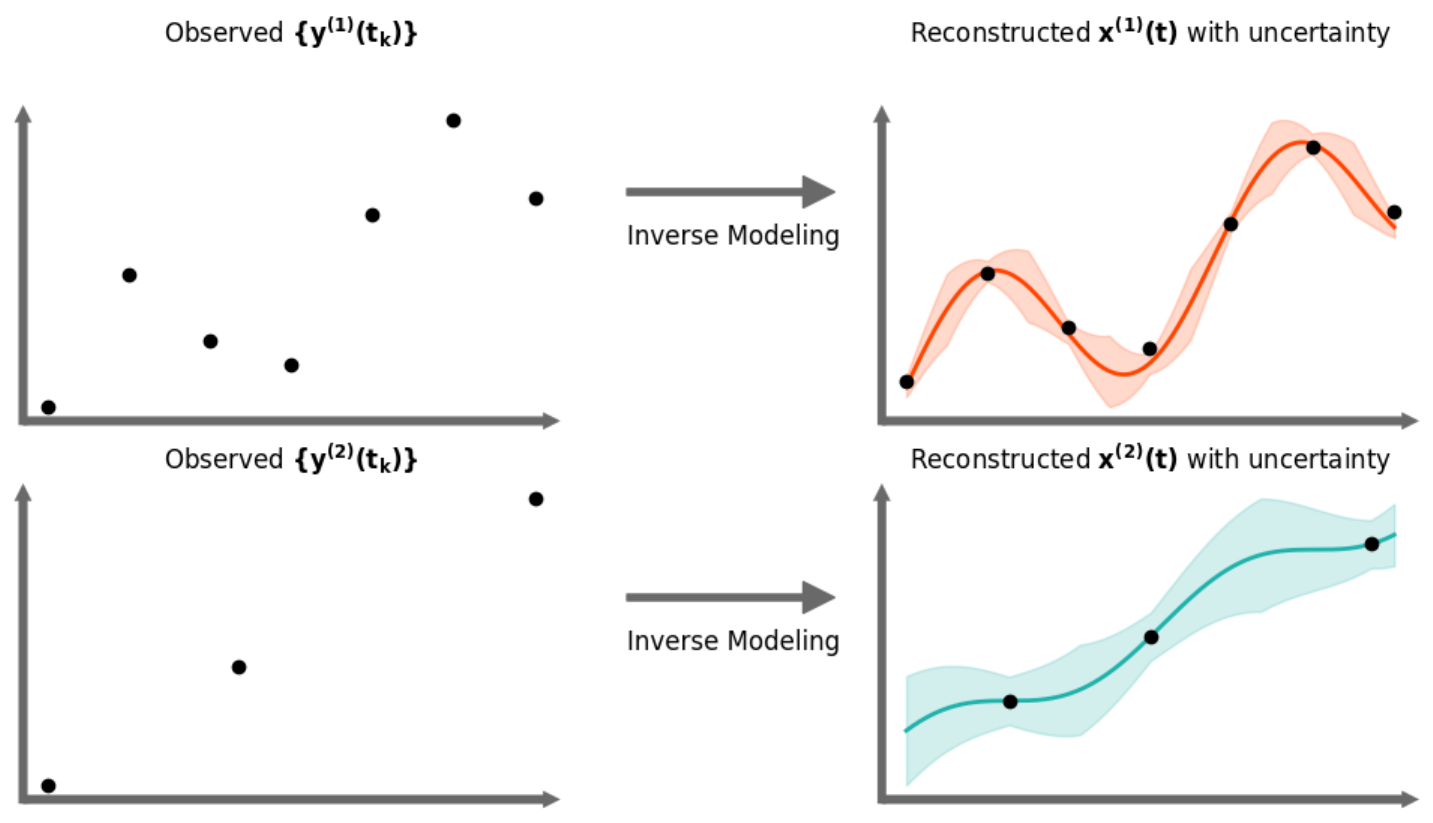}
\caption{Conceptual illustration of probabilistic interpolation as an inverse modeling task. Starting from sparse and noisy observations \( \{\mathbf{y}^{(i)}(t_k)\} \) (left), the process (model of choice) reconstructs the underlying continuous signal \( \mathbf{x}^{(i)}(t) \) (right). The indices \( (i) \) refer to different light curves. The shaded areas highlight the quantified uncertainty, demonstrating how the model addresses ambiguity in unobserved data regions. Crucially, information is shared across light curves: observations in one band are used to inform the interpolation of others, and vice versa.
}
\label{fig:interpolation_cartoon}
\end{figure*}

Despite significant advances, astrophysical time series data pose unique challenges (e.g., \citealt{hlozek_data_2019}). Observations are often irregularly sampled due to the constraints of telescope scheduling and weather conditions (\citealt{hlozek_data_2019}). Additionally, these data are frequently noisy, multivariate, and incomplete, making analysis difficult (e.g., \citealt{boyce_multiwavelength_2022}). The variability in astrophysical sources can also occur over different timescales, further complicating attempts to construct coherent models (e.g., \citealt{hlozek_data_2019}). Developing robust methods to impute (or interpolate) data is crucial for uncovering the detailed physics of these complex systems.

A compelling case study for addressing these challenges is Sagittarius A* (Sgr A*), the supermassive BH at the center of our galaxy. With a mass of approximately $4 \times 10^6 M_\odot$ and a distance of only $\sim 8.3$ kpc from Earth, Sgr A* is the closest supermassive BH (e.g., \citealt{gillessen_monitoring_2009, boehle_improved_2016, gillessen_update_2017, gravity_collaboration_constraining_2021, abuter_polarimetry_2023}). Despite having a low bolometric-to-Eddington luminosity ratio ($L/L_{Edd} = 10^{-9}$; e.g., \citealt{genzel_galactic_2010}), its proximity allows for high spatial and temporal resolution observations, which have provided critical insights into the dynamics of BHs and their immediate environments (e.g., \citealt{boyce_simultaneous_2019, boyce_multiwavelength_2022, collaboration_first_2022}). 

Multi-wavelength observations reveal that Sgr A* exhibits significant variability across all accessible wavelengths, including radio, infrared (IR), and X-rays (e.g., \citealt{boyce_simultaneous_2019, witzel_rapid_2021, boyce_multiwavelength_2022}). This variability is highly stochastic, with rapid flaring events superimposed on longer-term trends (e.g., \citealt{baganoff_chandra_2003, nowak_chandrahetgs_2012, neilsen_chandrahetgs_2013, witzel_near_2014, witzel_variability_2018, haggard_chandra_2019}). Additionally, the coordinated multi-wavelength campaigns often have limited temporal overlap, as logistical challenges make simultaneous observations difficult to sustain over extended periods (e.g., \citealt{eckart_coordinated_2008, boyce_simultaneous_2019, boyce_multiwavelength_2022}). These factors collectively result in sparse, irregularly sampled light curves that hinder direct multi-wavelength analyses of Sgr A*’s variability, emphasizing the need for reliable methods to reconstruct missing data (e.g., \citealt{boyce_multiwavelength_2022}).

Formally, the interpolation problem can be viewed as an instance of inverse modeling: we observe a sparse and noisy set of data points \( \{ \mathbf{y}(t_k) \} \), sampled from an underlying continuous process \( \mathbf{x}(t) \), and seek to reconstruct the most likely realizations of \( \mathbf{x}(t) \) consistent with these observations, as illustrated in Figure~\ref{fig:interpolation_cartoon} (e.g., \citealt{aster_parameter_2005, kaipio_statistical_2006}). Ideally, the model should also quantify uncertainty, reflecting the ambiguity inherent in unobserved regions of the light curve. This motivates the need for probabilistic models that can interpolate missing data while capturing complex temporal and inter-band dependencies (e.g., \citealt{paugnat_new_2024}).

One commonly used solution to this inverse problem is the Gaussian Process (GP), which models time series as continuous functions where each point is correlated with its neighbors according to a kernel (e.g., \citealt{rasmussen_gaussian_2008}). In essence, a GP assumes that the data lie on a curve with covariance properties specified by the kernel, and it uses the observed points to infer the likely shape of the missing regions.

GPs are particularly appealing in astrophysics because they provide probabilistic outputs and have a well-established theoretical foundation (e.g., \citealt{foreman-mackey_fast_2017, aigrain_gaussian_2023}). In particular, the study by \cite{paugnat_new_2024} demonstrates the practical effectiveness of multi-output GPs (MOGPs) in reconstructing light curves of Sgr A* observed at two interleaved near-infrared (NIR) wavelengths, enabling precise measurements of the spectral index despite the large flux variability. Their use of MOGPs illustrates how this class of models can capture correlated variability across wavelengths while providing robust uncertainty estimates, making them well-suited for sparse time series. However, despite these strengths, GPs scale poorly with the dimensionality of the data and the number of observations, limiting their practicality for the large, multivariate datasets increasingly encountered in modern astrophysics (e.g., \citealt{alvarez_computationally_2011, fagin_latent_2024}).

Related approaches in the machine learning literature, such as neural ODE-based models for irregularly sampled time series \citep{rubanova_latent_2019, kidger_neural_2020}, demonstrate the potential of continuous-time modeling. More recently, \citealt{fagin_latent_2024} proposed latent stochastic differential equations to model quasar variability, while \citealt{boersma_transformer_2024} applied transformers to model gamma-ray burst time series. Both approaches demonstrate strong predictive performance; however, the former relies on recurrent neural networks, which suffer from inefficient and unstable training (e.g., \citealt{pascanu_difficulty_2013}). The latter, while effective, does not offer probabilistic predictions in its current implementation, an essential feature for modeling the uncertainty inherent in Sgr A*'s light curves.

In this work, we introduce a diffusion-based generative framework for interpolating multivariate astrophysical time series with irregular sampling and missing values (e.g., \citealt{bilos_modeling_2023}). This model efficiently captures intricate temporal correlations across multiple bands while providing reliable uncertainty estimates. This enables robust reconstructions of sparse, multi-wavelength light curves and facilitates time-resolved studies of Sgr A*’s variability across the electromagnetic spectrum. We compare our approach against two minimal baselines: a MOGP model and a transformer-based model capable of predicting uncertainties (e.g., \citealt{alvarez_kernels_2012, yalavarthi_tripletformer_2023}).

Our approach builds on recent advances in the machine learning community, where significant efforts are being made to develop models that can handle irregular time series while providing meaningful probabilistic outputs (e.g. \citealt{wang_deep_2024}). This remains an active area of research, as similar challenges arise in various fields, such as medicine (e.g., \citealt{yalavarthi_tripletformer_2023}), finance (e.g., \citealt{gajamannage_real-time_2023}), and climatology (e.g., \citealt{lim_time-series_2021}). Despite the broad applicability, robust and scalable solutions are still difficult to achieve, particularly when dealing with complex data (e.g., \citealt{bilos_modeling_2023, ashok_tactis-2_2023, wang_deep_2024}). By applying cutting-edge techniques to the astrophysical domain, we not only seek to advance BH variability studies but also contribute to the development of general-purpose models for time series imputation.

This paper is structured as follows. In Section~\ref{sec:data}, we outline the datasets and preprocessing steps used for this study. Section~\ref{sec:model and methods} introduces the machine learning frameworks, describing the model architectures in detail. Section~\ref{sec:results-discussion} presents and discusses the results of our imputation experiments, including comparisons to existing methods and their implications for modeling Sgr A*’s variability. We conclude in Section~\ref{sec:conclusion} by outlining possible avenues for future improvement and extension of this work.

\section{Data Samples} \label{sec:data}

\begin{figure*}[ht]
\plotone{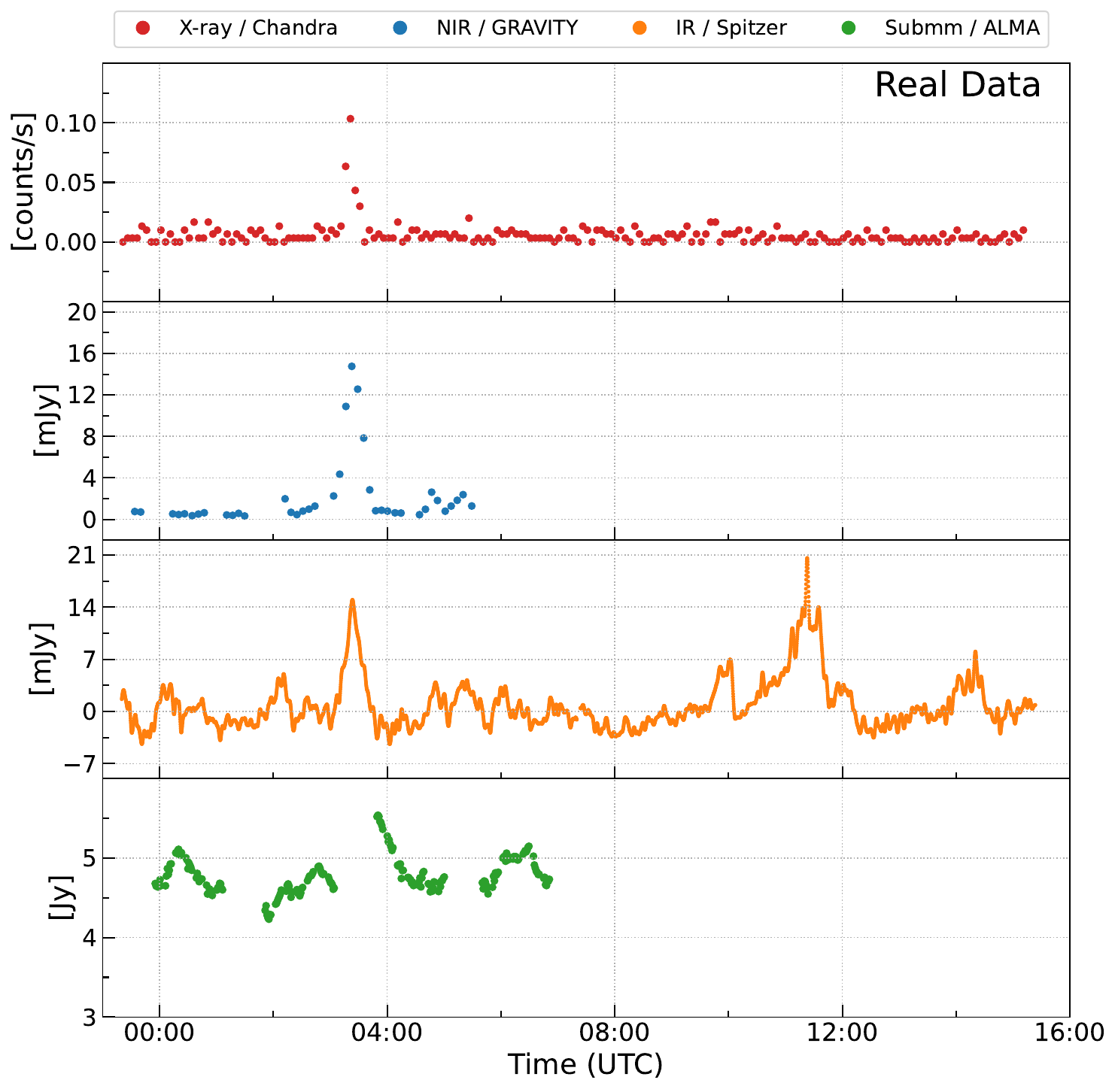}
\caption{Multi-wavelength observational data of Sgr~A* from the July 17–18, 2019 campaign, adapted from \citet{boyce_multiwavelength_2022}. The panels show flux measurements in four distinct bands: X-ray (Chandra, 2–8 keV, top), NIR (GRAVITY, 2.2 $\mu$m, second), IR (Spitzer, 4.5 $\mu$m, third), and submm (ALMA, 340 GHz, bottom). These data exhibit irregular sampling and varying noise levels across modalities, with dense temporal coverage in X-ray and IR bands and sparser, noisier measurements in NIR and submm.}
\label{fig:observationdata}
\end{figure*}

Our study relies on two sources of data: observational and simulated. The observational data serve as a case study, providing real-world measurements of Sgr A* across multiple wavelengths and enabling the qualitative evaluation of model outputs in a realistic setting. The simulated data, on the other hand, are used both to train our machine learning models and to quantitatively evaluate their performance. By generating synthetic light curves with known properties, we can rigorously assess each model’s ability to reconstruct missing or irregularly sampled data prior to applying it to real observations.

\subsection{Observational Data}
\label{subsec: observational data}

In July 2019, an observational campaign targeting Sgr A* was conducted using a coordinated suite of telescopes operating across the electromagnetic spectrum, including Chandra (2–8 keV; e.g., \citealt{baganoff_chandra_2003}), GRAVITY (2.2 $\mu$m; e.g., \citealt{gravity_collaboration_constraining_2021}), Spitzer (4.5 $\mu$m; e.g., \citealt{hora_spitzerirac_2014}), and ALMA (340 GHz; e.g., \citealt{michail_multiwavelength_2021}). Three epochs of simultaneous observations were acquired during the campaign, enabling a detailed, multi-wavelength view of Sgr A*’s variability. Figure~\ref{fig:observationdata} shows the July 17–18 epoch, which is the only one that includes GRAVITY coverage. Each instrument contributed complementary temporal and spectral information, but differences in cadence and scheduling constraints naturally introduced gaps and irregularities in the combined dataset. The data used in this work were originally compiled and reduced by \cite{boyce_multiwavelength_2022}. This subsection summarizes the reduction techniques employed and additional details can be found in the original publication.

Chandra observed Sgr A* in the 2–8 keV range using the ACIS-S3 chip in FAINT mode. Data were reduced with CIAO tools, correcting for background contamination and applying barycentric corrections (see \citealt{baganoff_chandra_2003} and \citealt{boyce_multiwavelength_2022} for more details). NIR observations from GRAVITY in the K-band (2.1--2.4~$\mu$m) provided high-resolution flux measurements, corrected for extinction and contamination from nearby sources such as S2 (see \citealt{gravity_collaboration_constraining_2021} and \citealt{boyce_multiwavelength_2022} for more details). An extinction correction of $A_K = 2.42$~mag was applied to the K-band light curves, following \cite{fritz_line_2011}. Simultaneously, Spitzer observations at 4.5~$\mu$m were conducted over three 16-hour epochs using the IRAC instrument in subarray mode. Light curves were extracted with corrections for intrapixel sensitivity variations and telescope jitter, and an offset of 1.9~mJy was applied based on prior studies (see \citealt{hora_spitzerirac_2014} and \citealt{boyce_multiwavelength_2022} for more details). For the Spitzer data, an extinction correction of $A_M = 0.97$~mag was used, also based on \cite{fritz_line_2011}. At millimeter wavelengths, ALMA observations were conducted using the 7~m compact array, covering the same epochs as Spitzer. Light curves were derived by fitting a point-source model to the visibilities, with absolute flux calibration accurate within 10\%. Heliocentric corrections were applied to align ALMA data with Spitzer (see \citealt{michail_multiwavelength_2021} and \citealt{boyce_multiwavelength_2022} for more details).

\subsection{Simulated Data}
\label{subsec:simulated data}

To provide a controlled environment for testing our interpolation methods, we use simulated light curves generated from the semi-empirical radiative model described in \cite{witzel_rapid_2021}. This model describes the variability of Sgr~A* across the electromagnetic spectrum as the result of two correlated stochastic processes, a fast and a slow component, each reflecting distinct physical mechanisms. The slow process captures variability in particle injection, governed by synchrotron cooling and particle escape, while the fast component reflects rapid changes in the high-energy electron population affecting the cutoff of the synchrotron spectrum (e.g., \citealt{ponti_powerful_2017}). The X-ray emission is produced via synchrotron self-Compton scattering of photons near the synchrotron peak, making it insensitive to the fast fluctuations and effectively acting as a low-pass filter on the NIR variability.

The model parameters are obtained through approximate Bayesian computation fitting to observed light curves, ensuring that the simulations reproduce key statistical properties of Sgr A*'s variability. These include the power spectra of X-ray, NIR, IR, and submillimeter (submm) light curves, the time lags between variability in different bands and the overall flux distributions across wavelengths. 

We generate a total of 16,350 simultaneous 24-hour light curves at 1-minute cadence across all four wavelengths. These are split into training (60\%), validation (30\%), and test (10\%) sets, corresponding to 9,810, 4,905, and 1,635 examples respectively (e.g., \citealt{goodfellow_deep_2016}). Each light curve is standardized to have zero mean and unit variance (e.g., \citealt{ivezic_statistics_2014}).

To mimic real observational constraints, we introduce a masking procedure that replicates the irregular and sparse cadences characteristic of the multi-wavelength observational campaigns. The masking strategy is customized for each band and is in concordance with the details in \cite{boyce_multiwavelength_2022}:

\begin{itemize}
    \item For the submm band, we simulate four observing blocks, each lasting 76 minutes and separated by 40-minute gaps. Within each epoch, observations occur in 7-minute scans spaced by 5-minute gaps, approximating submm telescope schedules.
    \item For the NIR band, we use a similar observing window as the submm band but introduce a temporal shift of up to ±50 minutes. Within this window, observations alternate between 1–10 minute integration periods and 1–10 minute gaps, mimicking sporadic ground-based access.
    \item For the IR and X-ray bands, we apply random subsampling by removing a fixed percentage of unmasked points (typically 30\%), simulating data loss due to reduced sensitivity.
    \item  We apply a 20\% random subsampling of unmasked data in each band to simulate additional data loss. While not physically motivated, this augmentation strategy is commonly used in time series imputation tasks (e.g., \citealt{choi_rdis_2023}) to prevent overfitting.
\end{itemize}

\begin{figure*}[ht]
\plotone{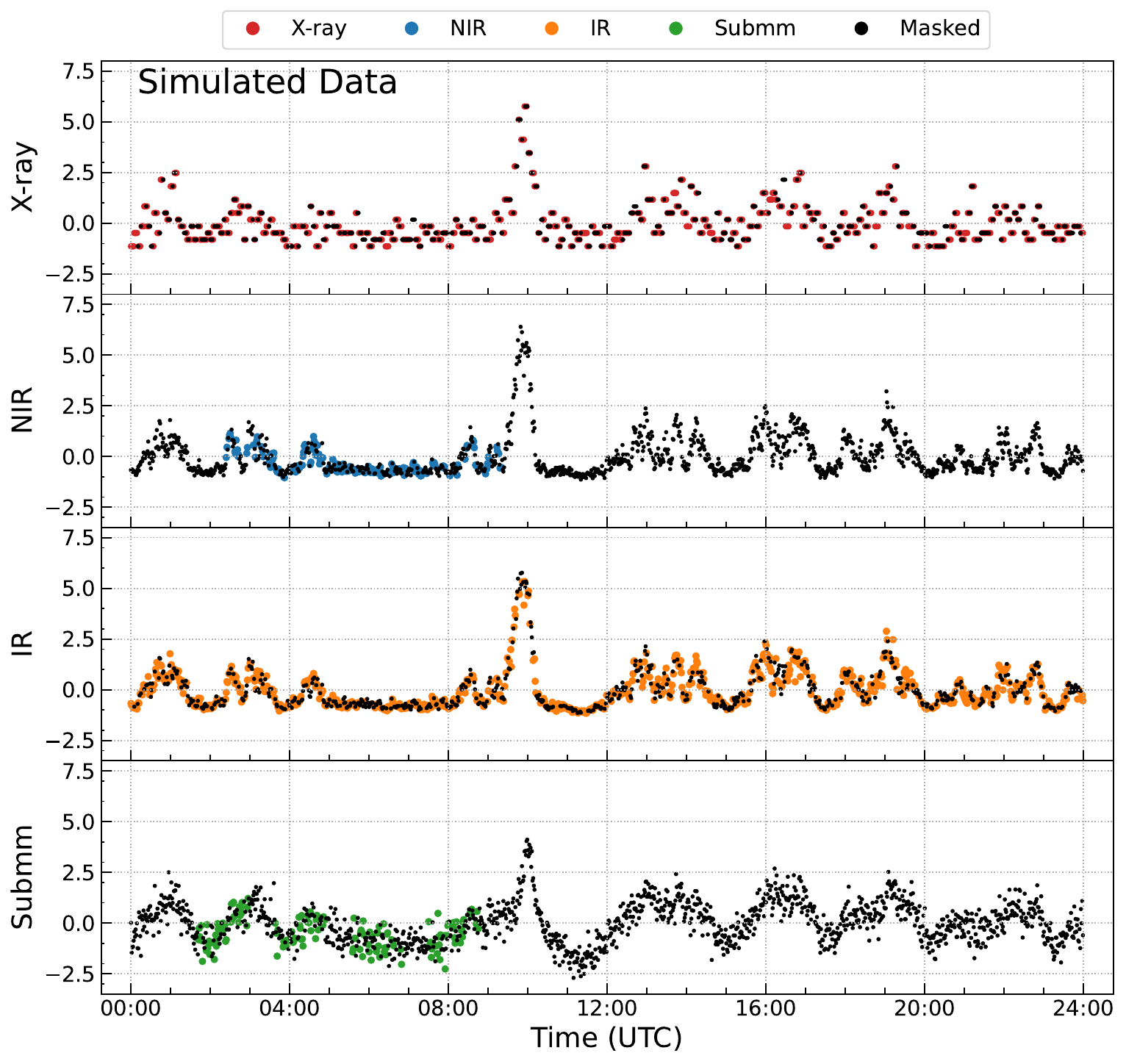}
\caption{
Example of simulated multi-wavelength light curves used for training and evaluation, drawn randomly from the test set. Each panel corresponds to one wavelength channel (X-ray, NIR, IR, submm), and displays a sample light curve generated to mimic the variability patterns, noise characteristics, and irregular sampling in real Sgr A* observations. The simulations are adapted from the framework described in \citet{witzel_rapid_2021}. A masking procedure is added to the simulated light curve to replicate telescope observation cadences. Colored circles represent observed (unmasked) data points and black circles represent masked data points.}
\label{fig:simulationdata}
\end{figure*}

An random example of a light curve from the test set after the masking procedure is shown in Figure~\ref{fig:simulationdata}. At each timestep, we record both the observed flux and a binary mask indicating whether the value was masked or not (e.g., \citealt{che_recurrent_2018, yalavarthi_tripletformer_2023}). This structure enables us to easily implement model training and inference.

By constructing a large ensemble of synthetic light curves with sampling gaps, we create a testbed for systematically assessing the performance and robustness of our interpolation models before applying them to real data (e.g., \citealt{cao_brits_2018}).

We note that all simulated light curves are generated in relative time, meaning that each realization is initialized at $t=0$ and subsequently masked according to the cadence patterns described above. This design choice allows us to test model robustness across arbitrary phase alignments and prevents the models from implicitly relying on absolute historical timestamps. While generating simulations in absolute time could better preserve the exact cadence and gap structure of past campaigns, our relative-time approach ensures that the interpolation methods generalize to a wide range of sampling configurations. This is particularly important given the irregular and campaign-specific nature of Sgr~A* observations, where no single cadence pattern is universally representative (e.g., \citealt{witzel_rapid_2021}).

\section{Models and Methods} \label{sec:model and methods}

Reconstructing missing observations in multi-wavelength time series is a fundamental inverse problem in time-domain astrophysics. Let \( \mathbf{x^{(i)}(t)} \) denote the underlying latent signals we wish to recover, and \( \mathbf{\{y^{(i)}(t_k)}\} \) the sparse and noisy sets of available observations. Our goal is to infer the distribution over plausible realizations of \( \mathbf{x^{(i)}(t)} \) conditioned on \( \mathbf{\{y^{(i)}(t_k)}\} \), while quantifying uncertainty (e.g., \citealt{aster_parameter_2005, kaipio_statistical_2006, rasmussen_gaussian_2008}). This process is illustrated in Figure \ref{fig:interpolation_cartoon}. This requires models that can reason over irregular time grids, capture cross-band correlations, and provide probabilistic outputs.

We explore three families of models for this task: MOGP regression (Section \ref{subsec:mogp}), diffusion models over continuous stochastic processes (Section \ref{subsec:cspd}), and transformer-based architectures (Section \ref{subsec:triplet}). We then define the evaluation metrics which are used to compare the model performances (Section \ref{subsec:evaluation-metrics}).

\subsection{Multi-Output Gaussian Process Regression}
\label{subsec:mogp}

A natural approach to modeling irregularly sampled time series is to treat each channel as a continuous function of time with a GP prior. GPs are flexible models that define distributions over functions and offer analytical uncertainty estimates (e.g., \citealt{rasmussen_gaussian_2008}).

We use MOGP regression as a baseline due to its interpretability and principled probabilistic framework. Unlike standard GPs, MOGPs enable the simultaneous inference of the posterior distributions of multiple variables while accounting for correlations between them (e.g., \citealt{bonilla_multi-task_2007, alvarez_kernels_2012}). This is done by assuming a GP prior over the latent functions \( f \), with zero mean:

\begin{equation}
    f(\mathbf{t}) \sim \mathcal{N}\left(0,\; K^f \otimes K^t + D \otimes I\right),
    \label{eq:ICM}
\end{equation}
where \( f(\mathbf{t}) \) is the joint vector of latent function values evaluated at observation times \( \mathbf{t} \), \( K^f \) is a similarity matrix (designed to capture correlation between processes), \( K^t \) is the covariance matrix computed over time points using a kernel function \( k_t(t, t') \), \( D \) is a diagonal matrix of observation noise per process, \( I \) is the identity matrix, and \( \otimes \) denotes the Kronecker product. This formulation is known as the intrinsic coregionalization model (ICM; e.g., \citealt{goovaerts_geostatistics_1997, alvarez_computationally_2011}).

We implement this model using the \texttt{GPyTorch} library (e.g., \citealt{gardner_gpytorch_2018}), which is built on the \texttt{PyTorch} framework (e.g., \citealt{paszke_automatic_2017}) and supports GPU acceleration, significantly enhancing computational efficiency. In our implementation, the temporal covariance matrix \( K^t \) is constructed using a radial basis function (RBF) kernel:

\begin{equation}
    k_t(t, t') = \exp\left(-\frac{(t - t')^2}{2\sigma^2}\right),
    \label{eq:rbfkernelequation}
\end{equation}
where \( \sigma \) is the kernel bandwidth. This kernel captures temporal smoothness, with closer time points being more strongly correlated. The task covariance matrix \( K^f \) is parameterized using a low-rank coregionalization structure, \( K^f = \mathbf{W}\mathbf{W}^\top \), where \( \mathbf{W} \) is a \( P \times R \) matrix (with \( P \) the number of wavelength bands and \( R=1 \) the rank), learned during training. This low-rank structure captures cross-band correlations while maintaining computational efficiency. The observation noise matrix \( D \) is implemented as a diagonal matrix with a shared noise variance parameter learned via a Gaussian likelihood (e.g., \citealt{gardner_gpytorch_2018}). The full covariance structure thus takes the form of Equation~\ref{eq:ICM}, with \( K^f \otimes K^t \) capturing the structured correlations and \( D \otimes I \) accounting for observation uncertainty.

GPs provide a minimal baseline because they make strong assumptions about the data's Gaussianity and correlation structure through the choice of kernel (e.g., \citealt{rasmussen_gaussian_2008}). To overcome this, we adopt a more flexible Bayesian perspective. Rather than assuming a fixed prior, we can learn complex distributions over functions from data using neural generative models (e.g., \citealt{dupont_generative_2022}). These models do not require a predefined basis and can express non-Gaussian, nonlinear behaviors more faithfully (e.g., \citealt{garnelo_conditional_2018}).

We build on recent advances in generative modeling to formulate interpolation as Bayesian inference over a learned function space (e.g., \citealt{dupont_generative_2022}). The model we use for this purpose is a diffusion model over continuous stochastic processes, described next (e.g., \citealt{bilos_modeling_2023}).

\subsection{Continuous Stochastic Process Diffusion}
\label{subsec:cspd}

Diffusion models define a generative process by progressively corrupting data with noise and then learning to reverse this process. The classical formulation (e.g., \citealt{song_score-based_2020}) applies to vector-valued data, where the signal is represented by a fixed-dimensional vector and noise is added independently across dimensions.

The continuous stochastic process diffusion (CSPD) model extends this framework to time series by treating the data as samples from a continuous stochastic process, such as a GP, rather than independent Gaussian vectors. This allows the model to account for temporal structure in the corruption process (e.g., \citealt{bilos_modeling_2023}).

We distinguish between \textbf{diffusion time} \( s \), an artificial time variable that parameterizes the forward and reverse diffusion process, and \textbf{light curve time} \( t \), which corresponds to the actual temporal axis of the time series being modeled.

The forward process is described by a stochastic differential equation:

\begin{equation}
d\mathbf{x}_s(t) = f(\mathbf{x}_s(t), s) \, ds + g(s) \, d\mathbf{w}_s(t),
\label{eq:forward_sde}
\end{equation}
where \(d\mathbf{x}_s(t) \) is the infinitesimal change in the signal at diffusion time \( s \), \( ds \) is the corresponding increment in time, \( f(\mathbf{x}_s(t), s) \) is a drift term, \( g(s) \) scales the injected noise, and \( d\mathbf{w}_s(t) \) is a standard Wiener process. The added noise is drawn from a GP:

\begin{equation}
\epsilon(t) \sim \mathcal{GP}(0, k(t_i, t_j)),
\label{eq:gpnoiseprocess}
\end{equation}
with a kernel \( k(t_i, t_j) = \exp(-\gamma (t_i - t_j)^2) \) ensuring temporal smoothness. The stochastic differential equation in Equation~\ref{eq:forward_sde} thus produces trajectories that evolve in both diffusion time \( s \) and physical time \( t \), with structured noise injected at each step. In this formulation, the kernel \( k(t_i, t_j) \) encodes only temporal correlations within each wavelength. We do not impose an explicit cross-band kernel, the multivariate dependencies are learned by the neural score network, which conditions on all bands jointly. This separates prior assumptions (temporal smoothness within bands) from data-driven learning of cross-band variability.

The reverse process learns to remove the added noise:

\begin{align}
d\mathbf{x}_s(t) &= \left[ f(\mathbf{x}_s(t), s) - g(s)^2 \nabla_{\mathbf{x}_s(t)} \log p_s(\mathbf{x}_s(t)) \right] ds \nonumber \\
&\quad + g(s) \, d\bar{\mathbf{w}}_s(t),
\label{eq:reverse_sde}
\end{align}
where \( \nabla_{\mathbf{x}_s(t)} \log p_s(\mathbf{x}_s(t)) \) is the score function and \( d\bar{\mathbf{w}}_s(t) \) is a reverse-time Wiener process. With a GP prior, the score function has a closed-form:

\begin{equation}
\nabla_{X_s} \log q(X_s \mid X_0) = - \tilde{\Sigma}^{-1} (X_s - \tilde{\mu}),
\end{equation}
where \( q(X_s \mid X_0) \) is the marginal distribution of the forward process at time \( s \) given the original data \( X_0 \), and \( \tilde{\mu} \) and \( \tilde{\Sigma} \) are the corresponding conditional mean and covariance. The neural network learns to predict the noise component, enabling efficient reparameterization.

Following \citet{bilos_modeling_2023}, we train the model by minimizing the difference between the predicted noise and the true noise injected during the forward process. The loss function takes the form:

\begin{equation}
\mathcal{L} = \mathbb{E}_{\epsilon, s} \left\| \epsilon_\theta\left(\sqrt{\bar{\alpha}_s} X_0 + \sqrt{1 - \bar{\alpha}_s} \, \epsilon, t, s \right) - \epsilon \right\|_2^2,
\end{equation}
where \( \epsilon_\theta \) is the neural network’s prediction of the noise, \( X_0 \) is the original time series, \( \epsilon \) is the sampled noise (see Equation \ref{eq:gpnoiseprocess}), \( t \) is the observation time grid, \( s \) is the index corresponding to the sampled diffusion step, and \( \bar{\alpha}_s \) is the cumulative noise schedule at step \( s \). This formulation allows training without directly computing the full GP covariance matrix, improving computational efficiency.

While traditional models often project the data onto a predefined analytic basis (e.g., Fourier or wavelets), CSPD instead uses a learnable, data-driven projection based on similarity kernels. Following \citet{bilos_modeling_2023}, the model applies a learned RBF kernel to compute similarities between observed and query time points. Specifically, the observed inputs \( \mathbf{\{y^{(i)}(t_k)}\} \) at their respective time coordinates are first encoded into a latent representation using a neural network. This representation is then projected to the query times via kernel multiplication, enabling information transfer from observed to unobserved regions of the time series. This mechanism functions as a flexible and permutation-invariant projection scheme, tailored for irregular sampling. This means that during training and inference, the model is provided with both the observation times and a binary mask indicating which entries are observed. 

At inference, conditioning is enforced by fixing observed values at each diffusion step and updating only the unobserved entries. Specifically, the reverse-time update at step \( s \) takes the form (e.g., \citealt{bilos_modeling_2023}):

\begin{equation}
\begin{split}
\mathbf{x}_{s-1}(t) &= \mathbf{m}_{\text{obs}} \odot \mathbf{y}_{\text{obs}}(t) + \mathbf{m}_{\text{miss}} \odot  \biggl[ \frac{1}{\sqrt{\bar{\alpha}_s}} \biggl( \mathbf{x}_s(t) -\\
&\quad \frac{1 - \bar{\alpha}_s}{\sqrt{1 - \bar{\alpha}_s}} \epsilon_\theta(\mathbf{x}_s(t), \mathbf{y}_{\text{obs}}(t), t, s) \biggr) + \sigma_s \epsilon_s(t) \biggr],
\label{eq:conditional_reverse}
\end{split}
\end{equation}
where \( \mathbf{m}_{\text{obs}} \) and \( \mathbf{m}_{\text{miss}} = 1 - \mathbf{m}_{\text{obs}} \) are binary masks indicating observed and missing locations, respectively, \( \odot \) denotes element-wise multiplication, \( \mathbf{y}_{\text{obs}}(t) \) are the fixed observed values from \( \mathbf{\{y^{(i)}(t_k)\}} \), \( \sigma_s \) is the step-dependent noise scale, and \( \epsilon_s (t) \) is the sampled noise from Equation~\ref{eq:gpnoiseprocess}. This ensures that observed values remain clamped to their true values throughout the reverse process, while missing entries are progressively denoised according to the learned noise \( \epsilon_\theta \), which conditions on both the noisy state \( \mathbf{x}_s(t) \) and the observed values \( \mathbf{y}_{\text{obs}}(t) \).

By modeling time series as continuous stochastic processes, CSPD ensures temporal consistency in generated samples, making it well-suited for interpolation tasks in irregularly sampled data (e.g., \citealt{rasul_autoregressive_2021, bilos_modeling_2023}).

\subsection{TripletFormer}
\label{subsec:triplet}

While diffusion models offer high flexibility, their sampling procedures are computationally intensive. Transformer models, by contrast, offer fast inference and strong performance on sequential data, including irregular time series (e.g., \citealt{vaswani_attention_2017, yalavarthi_tripletformer_2023}).

Transformers use an encoder-decoder architecture with attention mechanisms that enable the model to weigh the importance of different parts of the sequence when making predictions (e.g., \citealt{vaswani_attention_2017}). The attention mechanism forms the backbone of transformers and is mathematically defined as:

\begin{equation}
\text{Attention}(Q, K, V) = \text{softmax}\left(\frac{QK^\top}{\sqrt{d_k}}\right)V,
\end{equation}
where \( Q \) (query), \( K \) (key), and \( V \) (value) are learned linear projections of the input, and \( d_k \) is the dimensionality of the key vectors. The multi-head mechanism runs this operation in parallel across multiple attention heads, enabling the model to attend to different aspects of the sequence simultaneously (e.g., \citealt{vaswani_attention_2017}).

By leveraging attention, the model learns both inter-channel correlations (relationships between different variables in a multivariate time series) and intra-channel correlations (patterns within each variable over time; e.g., \citealt{yalavarthi_tripletformer_2023}).

\begin{figure}[ht]
\plotone{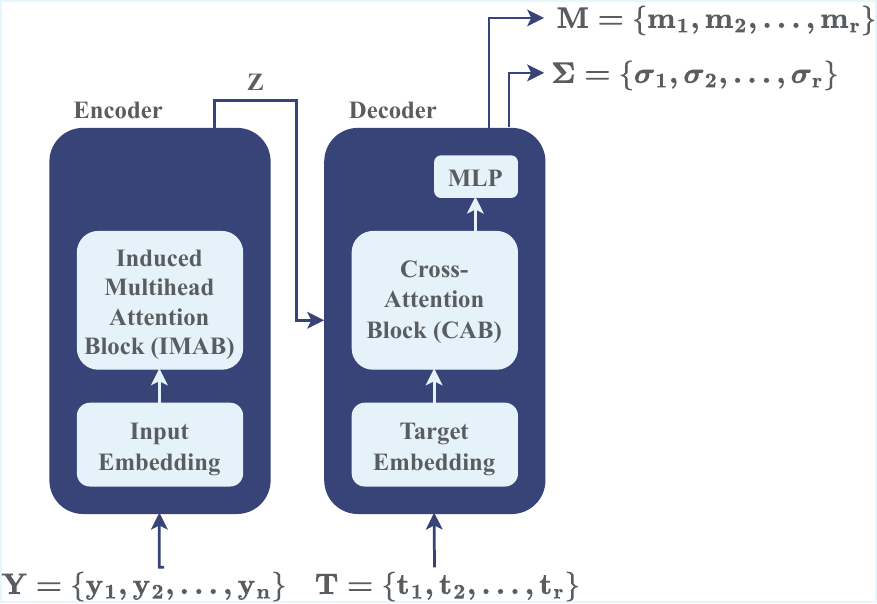}
\caption{
Architecture of the TripletFormer model (\citealt{yalavarthi_tripletformer_2023}). The encoder processes observed inputs via attention to produce a latent representation \( Z \). The decoder uses cross-attention with target queries to produce predictive means \( \mu \) and standard deviations \( \sigma \), defining a Gaussian distribution over missing values.
}
\label{fig:tripletformer}
\end{figure}

We implement our transformer model using the TripletFormer architecture proposed by \citet{yalavarthi_tripletformer_2023}, shown in Figure~\ref{fig:tripletformer}. This architecture consists of an encoder with an input embedding layer and an induced multi-head attention block, that outputs a latent representation $Z$. The decoder comprises a target embedding layer, a multi-head cross-attention block, and a feed-forward layer. The final layer of the decoder outputs the parameters of a Gaussian distribution for each target point, representing the predictive uncertainty over the missing values.

The model is trained using a heteroscedastic loss function that combines the negative log-likelihood of the predicted Gaussian distribution with a mean squared error (MSE) regularization term. The loss function is defined as (e.g., \citealt{yalavarthi_tripletformer_2023}):

\begin{equation}
\mathcal{L} = \sum_{n=1}^{N} \left[ - \mathbb{E} \left[ \log \hat{P}(U'_n \mid Y_n, T_n) \right] + \lambda \left\| U'_n - M_n \right\|_2^2 \right],
\end{equation}
where \( U'_n \) is the ground truth value at the \( n \)-th query time step, \( Y_n \) is the set of observed context points, \( T_n \) is the set of query times, \( \hat{P}(U'_n \mid Y_n, T_n) \) is the predicted Gaussian distribution, \( M_n \) is the predicted mean, and \( \lambda \) is a weighting factor balancing the negative log-likelihood and MSE terms. The negative log-likelihood encourages accurate uncertainty estimates, while the MSE term prevents the model from collapsing to flat mean predictions early during training (e.g., \citealt{yalavarthi_tripletformer_2023}).

\subsection{Model Evaluation Metrics} \label{subsec:evaluation-metrics}

To assess the quality of the reconstructed light curves, we use two complementary metrics: the MSE and the continuous ranked probability score (CRPS). These metrics allow us to evaluate both the accuracy of the mean predictions and the quality of the predicted uncertainty (e.g., \citealt{hersbach_decomposition_2000, gneiting_strictly_2007}).

 The MSE is defined as:
\begin{equation}
    \text{MSE} = \frac{1}{N} \sum_{i=1}^{N} (y_i - \hat{y}_i)^2,
\end{equation}
where \( y_i \) denotes the true flux value at time step \( i \), \( \hat{y}_i \) is the model's predicted mean at that time step, and \( N \) is the total number of data points in the test set. MSE provides a pointwise evaluation of the prediction error, without accounting for uncertainty (e.g., \citealt{bishop_pattern_2006}).

CRPS evaluates the quality of a probabilistic forecast by comparing the predicted cumulative distribution function, \( F(z) \), to the empirical cumulative distribution function of the true observation \( y \). Formally, it is defined as (e.g., \citealt{matheson_scoring_1976, gneiting_strictly_2007}):

\begin{equation}
    \text{CRPS}(F, y) = \int_{-\infty}^{\infty} \left( F(z) - \mathbbm{1}\{z \geq y\} \right)^2 \, dz,
\end{equation}
where \( F(z) \) is the predicted cumulative distribution function evaluated at threshold \( z \), \( y \) is the observed value, and \( \mathbbm{1}\{z \geq y\} \) is the indicator function representing the empirical cumulative distribution function of \( y \), which equals 1 if \( z \geq y \) and 0 otherwise.

For models that output a Gaussian predictive distribution (e.g., MOGP and TripletFormer), the CRPS can be computed using a closed-form solution (e.g., \citealt{gneiting_strictly_2007}):

\begin{align}
\text{CRPS}(\mu, \sigma; y) =\; & \sigma \Bigg[ \frac{y - \mu}{\sigma} \left( 
2\Phi\left( \frac{y - \mu}{\sigma} \right) - 1 \right) \notag \\
& + 2\phi\left( \frac{y - \mu}{\sigma} \right) - \frac{1}{\sqrt{\pi}} \Bigg],
\end{align}
where \( \mu \) and \( \sigma \) are the predicted mean and standard deviation, and \( \Phi \), \( \phi \) are the cumulative distribution function and probability density function of the standard normal distribution, respectively.

For the CSPD model, which does not yield an analytical form of the predictive distribution, we use a quantile-based approximation. Specifically, we compute CRPS as the average quantile loss over a set of quantile levels (e.g., \citealt{marchesoni-acland_crps_2024}):
\begin{equation}
    \text{CRPS} \approx \frac{1}{K} \sum_{k=1}^{K} \frac{1}{Z} \cdot \mathcal{L}_{\text{quantile}}(q_k),
\end{equation}
where \( \mathcal{L}_{\text{quantile}}(q_k) \) is the quantile loss at level \( q_k \), \( K \) is the number of quantiles (typically 19 for quantiles between 0.05 and 0.95), and \( Z \) is a normalization factor to account for missing or masked data points. This approach enables a tractable and differentiable CRPS computation even when the predictive distribution is represented by samples and is validated in \cite{marchesoni-acland_crps_2024}.

\begin{figure*}[ht]
\includegraphics[width=\textwidth]{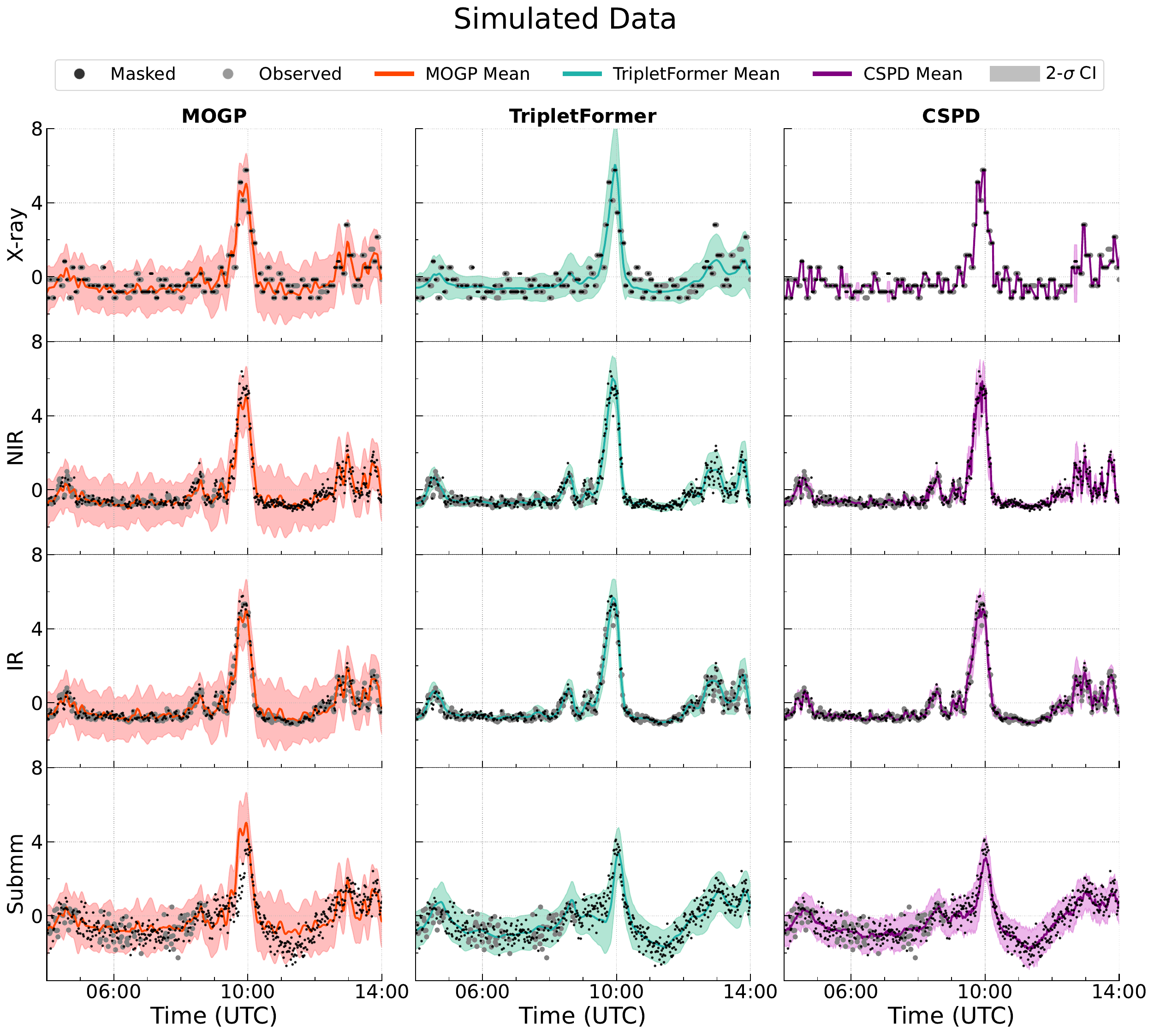}
\caption{Model interpolation results on the simulated light curve of Sgr A* from Figure~\ref{fig:simulationdata}, plotted from 4:00 UTC to 14:00 UTC for clarity. Each column corresponds to a different model (MOGP, TripletFormer, CSPD), and each row represents a different wavelength band (X-ray, NIR, IR, submm). Observed (unmasked) data upon which the models are conditioned are shown as grey circles, while test (masked) data points are shown as black circles. Solid lines denote the mean predictions of each model while shaded regions denote 2$\sigma$ credible intervals. All models successfully recover the underlying signal, but the CSPD model captures sharper variability and exhibits tighter uncertainty bounds compared to the smoother, more conservative predictions of the MOGP and TripletFormer. See Figure~\ref{fig:simulated data interpolation} for the full 24 hour interpolation.}
\label{fig:simulatedinterpolation per model}
\end{figure*}

\section{Results and Discussion} \label{sec:results-discussion}

In this section, we evaluate the performance of our proposed CSPD model in comparison to two established baselines: a MOGP model and the TripletFormer. We assess model performance based on both simulated and real Sgr A* light curves, focusing on reconstruction accuracy and the calibration of predictive uncertainties.

\subsection{Model Performances on Simulated Data} \label{subsec:simulated performances}

We begin by training the TripletFormer and CSPD models on a dataset of 9,810 simulated multi-wavelength light curves. These light curves are designed to mimic the variability and sampling characteristics typical of real Sgr A* observations, based on certain model assumptions because the true nature of correlations in the real data remains uncertain. Model performance is optimized using a separate validation set of 4,905 examples, with hyperparameters selected to minimize the validation loss. The results of the hyperparameter tuning is summarized in Appendix \ref{appendix:hyperparameter}, with Tables~\ref{tab:hyperparameters triplet} and~\ref{tab:hyperparameters cspd}, where the configuration highlighted in bold corresponds to the final model used for all subsequent evaluations.

After training, we assess model performance on a reserved test set of 1,635 simulated light curves. This allows us to evaluate the reconstruction capabilities of each approach under controlled conditions with known ground truth. We include a MOGP baseline, as described in Section~\ref{subsec:mogp}, which provides a principled probabilistic benchmark already used in astrophysical time series modeling. MOGPs have been previously adopted as baselines in recent machine learning applications to astrophysics, including the latent stochastic differential equation framework proposed by \citet{fagin_latent_2024}, where they served as a reference point for evaluating learned time series representations of quasar variability. They also feature prominently in the analysis by \citet{paugnat_new_2024}, where MOGPs were successfully applied to reconstruct interleaved NIR light curves of Sgr~A* for spectral index analysis. This highlights their continued relevance as a reliable baseline for handling multi-wavelength variability in sparse astronomical data.

Figure~\ref{fig:simulatedinterpolation per model} shows a representative example of model interpolation on the same test sample shown in Figure~\ref{fig:simulationdata}, plotted from 4:00 UTC to 14:00 UTC for clarity (See Figure~\ref{fig:simulated data interpolation} for full 24 hour interpolation). Qualitatively, we observe that all three models produce comparable interpolations that generally align well with the test data across all wavelength bands. Each model successfully captures the underlying structure of the light curves, demonstrating a strong ability to infer missing values from limited context.


\begin{deluxetable*}{ccccc}[ht]
\tablecaption{Model Comparison on Simulated Light Curves: MSE and CRPS by Wavelength} \label{tab:model performances}
\tablewidth{0pt}
\tablehead{
 \colhead{\textbf{Wavelength}} & 
\colhead{\textbf{Metric}} & \colhead{\textbf{MOGP}} & \colhead{\textbf{TripletFormer}} & \colhead{\textbf{CSPD}} 
}
\startdata
X-ray & MSE  & $0.169 \pm 0.082$ & $0.217 \pm 0.131$ & $0.011 \pm 0.003$ \\
X-ray & CRPS  & $0.234 \pm 0.047$ & $0.230 \pm 0.075$ & $0.021 \pm 0.002$ \\
NIR & MSE     & $0.116 \pm 0.048$ & $0.081 \pm 0.041$ & $0.026 \pm 0.004$ \\
NIR & CRPS    & $0.207 \pm 0.033$ & $0.126 \pm 0.034$ & $0.112 \pm 0.008$ \\
IR& MSE      & $0.091 \pm 0.040$ & $0.060 \pm 0.033$ & $0.026 \pm 0.005$ \\
IR & CRPS     & $0.195 \pm 0.031$ & $0.107 \pm 0.031$ & $0.101 \pm 0.008$ \\
Submm & MSE   & $0.690 \pm 0.283$ & $0.263 \pm 0.109$ & $0.186 \pm 0.017$ \\
Submm & CRPS  & $0.439 \pm 0.084$ & $0.284 \pm 0.056$ & $0.320 \pm 0.015$ \\
\hline
\textbf{Total Avg} & MSE  & $0.266 \pm 0.289$ & $0.155 \pm 0.079$ & $0.062 \pm 0.072$ \\
\textbf{Total Avg} & CRPS & $0.269 \pm 0.113$ & $0.186 \pm 0.049$ & $0.138 \pm 0.111$ \\
\enddata
\end{deluxetable*}

Notable differences, however, emerge in the predicted uncertainties. The MOGP model consistently exhibits broader credible intervals across all wavelengths, reflecting its conservative estimation of uncertainty (e.g., \citealt{bonilla_multi-task_2007}). In contrast, both the transformer-based and diffusion-based models yield narrower uncertainty bounds, suggesting greater confidence in their reconstructions. This difference can be attributed in part to the underlying modeling assumptions: MOGPs explicitly model uncertainty through a joint Gaussian prior, and depending on the choice of kernel, may produce smoother and more conservative predictions (e.g., \citealt{bonilla_multi-task_2007, rasmussen_gaussian_2008}). The TripletFormer, while capable of estimating uncertainties, is still constrained by its assumed parametric output distribution, a Gaussian (e.g., \citealt{yalavarthi_tripletformer_2023}). The CSPD diffusion model, by contrast, generates samples directly, inherently learning to model the full conditional distribution of the data, enabling sharper transitions and more localized variability in the reconstructed curves (e.g., \citealt{rasul_autoregressive_2021, bilos_modeling_2023}).

Another key observation is the consistent increase in uncertainty in the submm band across all models, meaning that the predicted credible intervals for this band are wider, indicating greater uncertainty about the true flux values. The wider credible intervals in this band reflect the greater intrinsic noise and lower sampling density typically associated with submm observations in real datasets (e.g., \citealt{marrone_submillimeter_2006}, \citealt{boyce_multiwavelength_2022}). In the simulations, this is manifested through increased scatter and sparser unmasked data for the submm channel, leading all models to express reduced confidence in their predictions for this band (e.g., \citealt{herrnstein_variability_2004, wielgus_millimeter_2022}). This behavior aligns with expectations and demonstrates that the models are sensitive to modality-specific data quality (e.g., \citealt{wilson_deep_2016}).

In contrast to the other wavelengths, the X-ray light curves present a fundamentally different statistical profile, shaped by their discrete and Poissonian nature. Unlike the smoother and continuous flux variations typically observed in the submm or IR, X-ray data is characterized by abrupt and sparse photon events (e.g., \citealt{witzel_rapid_2021}). This distinction is especially apparent in Figure~\ref{fig:simulatedinterpolation per model}, where only the CSPD model captures the qualitative aspect of X-ray variability. While both the MOGP and TripletFormer models tend to smooth over this structure, likely due to strong conditioning on correlated trends in other wavelengths, the CSPD model preserves the granular and bursty character of X-ray emission. This suggests that the CSPD model implicitly learns to disentangle modality-specific features and represent them with greater fidelity, rather than applying a uniform smoothness assumption across all wavelengths.

Quantitative performance metrics on the simulated test set are presented in Table~\ref{tab:model performances}, which reports both the MSE and CRPS for each model across individual wavelength bands, along with their overall averages. For both metrics, lower values indicate better performance, corresponding to more accurate point predictions (MSE) and better-calibrated uncertainty estimates (CRPS).

Across all wavelengths, the CSPD model achieves the lowest average error, with an MSE of 0.0621 and a CRPS of 0.1384. Its performance is particularly strong in the X-ray and NIR bands, where it outperforms both the MOGP and TripletFormer models by a substantial margin. These results confirm the model’s ability to capture fine-grained temporal variability, consistent with the sharper reconstructions observed in Figure~\ref{fig:simulatedinterpolation per model}.

The TripletFormer exhibits competitive performance in the NIR, IR, and submm bands, achieving better MSE and CRPS scores than the MOGP in all but the X-ray band. While it does not match the CSPD in overall accuracy, it nonetheless represents a strong baseline, especially given its scalability and capacity for probabilistic forecasting (e.g., \citealt{yalavarthi_tripletformer_2023}). Its relatively low CRPS values suggest that the model's uncertainty estimates are well-calibrated in most bands for the simulated data (e.g., \citealt{gneiting_strictly_2007}). However, this remains a preliminary indication, as its uncertainty calibration on real observational data will be assessed separately in Section~\ref{subsec:Uncertainty Calibration}.

The MOGP model performs the most poorly compared to all other models, with the highest average MSE (0.2664) and CRPS (0.2690). Its largest errors appear in the submm band, where the MSE reaches 0.6903, more than twice that of the transformer and nearly four times that of the diffusion model. This behavior likely stems in part from our choice of a RBF kernel, which is known to produce overly smooth interpolants and may under-represent rapid variability in the signal (e.g., \citealt{bonilla_multi-task_2007, rasmussen_gaussian_2008}). While this kernel is effective for capturing long-term trends, it can lead to conservative estimates in high-noise or data-sparse regimes, where sharp transitions are smoothed out. Similar limitations of single-kernel approaches in astrophysical time series modeling have been noted by \cite{paugnat_new_2024}, who show that combining multiple kernels, such as a RBF kernel for long-term structure and an exponential kernel for short-timescale variability, can offer a more flexible and expressive representation of multi-timescale dynamics.

Another contributing factor to the MOGP's underperformance may lie in its treatment of observational noise. In our implementation, the noise term is modeled as a Kronecker product \( D \otimes I \), where \( D \) is a diagonal matrix specifying a fixed noise level per wavelength and \( I \) is the identity matrix. This formulation assumes homoscedasticity, that is, constant noise variance across all time steps within each wavelength band (e.g., \citealt{alvarez_computationally_2011}). While this assumption simplifies inference and is commonly adopted in MOGP models, it neglects potential temporal variation in the observational uncertainties. In practice, measurement noise can vary due to instrumental effects, changing observing conditions, or flux-dependent signal-to-noise ratios (e.g., \citealt{boyce_multiwavelength_2022}). Incorporating heteroscedastic noise models that allow the variance to evolve over time could help better capture these effects and yield improved uncertainty quantification in future applications.

Despite these limitations, MOGPs have proven highly effective in astrophysical time series modeling when their architecture is carefully adapted to the problem. For instance, \citet{paugnat_new_2024} applied MOGPs to derive the spectral index of Sgr A* across a broad flux range, using them to interpolate interleaved NIR observations and constrain physical emission models. This example demonstrate that while the default MOGP implementation underperforms in this study, its principled probabilistic foundation remains a powerful tool when properly tailored to the specific complexities of the problem through the choice of the kernel (e.g., \citealt{paugnat_new_2024}).

In conclusion, the submm band yields the highest prediction error and uncertainty accross all models, consistent with the qualitative observations discussed earlier. The elevated MSE and CRPS values in this channel highlight the difficulty of reconstructing low signal-to-noise ratio, sparsely sampled observations, and the importance of developing models that can generalize well across variable data quality conditions (e.g., \citealt{aigrain_gaussian_2023}).

\subsection{Robustness to Missing Data} \label{subsec:missing data}

\begin{deluxetable}{cccc}[ht]
\tablecaption{MSE Across Varying Levels of Missing Data for Random and Burst Missing Tests} \label{tab:mse_missing}
\tablewidth{0pt}
\tablehead{
\colhead{\% of Missing Data} & \colhead{MOGP} & \colhead{TripletFormer} & \colhead{CSPD}
}
\startdata
\multicolumn{4}{c}{\textbf{Random Missing Data}} \\
\hline
10 & $0.264 \pm 0.291$ & $0.170 \pm 0.084$ & $0.070 \pm 0.092$ \\
30 & $0.268 \pm 0.286$ & $0.163 \pm 0.078$ & $0.074 \pm 0.083$ \\
50 & $0.273 \pm 0.280$ & $0.191 \pm 0.087$ & $0.086 \pm 0.076$ \\
70 & $0.285 \pm 0.267$ & $0.188 \pm 0.087$ & $0.123 \pm 0.075$ \\
90 & $0.354 \pm 0.251$ & $0.238 \pm 0.101$ & $0.287 \pm 0.053$ \\
95 & $0.467 \pm 0.239$ & $0.324 \pm 0.126$ & $0.387 \pm 0.065$ \\
\hline
\multicolumn{4}{c}{\textbf{Burst Missing Data}} \\
\hline
10 & $0.341 \pm 0.522$ & $1.072 \pm 1.260$ & $0.113 \pm 0.108$ \\
30 & $0.324 \pm 0.371$ & $0.287 \pm 0.191$ & $0.114 \pm 0.108$  \\
50 & $0.718 \pm 0.389$ & $5.807 \pm 1.420$ & $0.618 \pm 0.098$ \\
70 & $0.903 \pm 0.299$ & $9.359 \pm 1.822$ & $0.841 \pm 0.084$ \\
90 & $1.084 \pm 0.195$ & $2.787 \pm 0.851$ & $1.060 \pm 0.052$ \\
95 & $1.183 \pm 0.343$  & $6.142 \pm 1.429$ & $1.073 \pm 0.042$ \\
\enddata
\end{deluxetable}

\begin{deluxetable}{cccc}[ht]
\tablecaption{CRPS Across Varying Levels of Missing Data for Random and Burst Missing Tests} \label{tab:crps_missing}
\tablewidth{0pt}
\tablehead{
\colhead{\% of Missing Data} & \colhead{MOGP} & \colhead{TripletFormer} & \colhead{CSPD}
}
\startdata
\multicolumn{4}{c}{\textbf{Random Missing Data}} \\
\hline
10 & $0.268 \pm 0.113$ & $0.195 \pm 0.051$ & $0.144 \pm 0.127$ \\
30 & $0.269 \pm 0.113$ & $0.190 \pm 0.049$ & $0.153 \pm 0.115$ \\
50 & $0.270 \pm 0.113$ & $0.209 \pm 0.051$ & $0.170 \pm 0.103$ \\
70 & $0.266 \pm 0.122$ & $0.207 \pm 0.052$ & $0.211 \pm 0.090$ \\
90 & $0.300 \pm 0.110$ & $0.232 \pm 0.054$ & $0.347 \pm 0.057$ \\
95 & $0.342 \pm 0.096$ & $0.270 \pm 0.057$ & $0.412 \pm 0.058$ \\
\hline
\multicolumn{4}{c}{\textbf{Burst Missing Data}} \\
\hline
10 & $0.305 \pm 0.154$ & $0.540 \pm 0.313$ & $0.213 \pm 0.142$ \\
30 & $0.294 \pm 0.125$ & $0.256 \pm 0.082$ & $0.211 \pm 0.138$ \\
50 & $0.422 \pm 0.099$ & $1.772 \pm 0.260$ & $0.520 \pm 0.056$  \\
70 & $0.479 \pm 0.077$ & $2.155 \pm 0.231$ & $0.638 \pm 0.034$ \\
90 & $0.537 \pm 0.077$ & $1.178 \pm 0.240$ & $0.759 \pm 0.022$ \\
95 & $0.567 \pm 0.115$ & $1.823 \pm 0.167$ & $0.741 \pm 0.027$ \\
\enddata
\end{deluxetable}

To evaluate the robustness of each model to varying levels and structures of missing data, we conduct an additional set of experiments in which subsets of observations are masked according to two schemes: (\textit{i}) random missingness, where individual points are removed at increasing proportions (10\%, 30\%, 50\%, 70\%, 90\%, and 95\%), and (\textit{ii}) burst missingness, where contiguous segments are masked to mimic more realistic observational gaps. Each model is then used to interpolate the missing values based on the remaining context points. Tables~\ref{tab:mse_missing} and \ref{tab:crps_missing} present the resulting MSE and CRPS, averaged across all wavelengths and all test samples, for each level of data sparsity. This type of stress-testing procedure has been widely adopted in prior work on time series imputation (e.g., \citealt{cao_brits_2018, luo_multivariate_2018, bilos_modeling_2023}), as a means to assess model robustness under increasing uncertainty and structured data loss.

Across both tests and both metrics, all models show a clear degradation in performance as the fraction of missing data increases. Overall, CSPD achieves the best accuracy, maintaining the lowest MSE and CRPS across nearly all sparsity levels. At extreme sparsity (90\%–95\%) in the random missing test, however, CSPD is slightly outperformed by the TripletFormer. This suggests that while CSPD excels in moderately sparse settings, its reliance on iterative sampling may limit performance when the available context is extremely limited (e.g., \citealt{rasul_autoregressive_2021, bilos_modeling_2023}). The TripletFormer, by contrast, shows more stable degradation beyond 70\%, which may reflect its architectural advantage in capturing long-range dependencies from minimal context (e.g., \citealt{yalavarthi_tripletformer_2023}).

While competitive under random missingness, the TripletFormer, exhibits a sharp decline in both MSE and CRPS under burst missingness, with errors increasing by nearly an order of magnitude beyond 50\% sparsity. This sensitivity highlights the model’s reliance on continuous temporal context (e.g., \citealt{yalavarthi_tripletformer_2023}). The MOGP, although generally less accurate, remains very stable and consistent model across both missingness types and metrics. Its relative robustness to burst missingness suggests that Gaussian process priors provide smooth extrapolations across large gaps, allowing it to outperform the TripletFormer at high sparsity levels. This aligns with previous findings that GPs can offer consistent performance on short incomplete light curves that probe BH variability (e.g., \citealt{paugnat_new_2024, fagin_latent_2024}). These results collectively emphasize the importance of evaluating interpolation models under both random and structured data loss, as real astrophysical light curves often contain long, contiguous observational gaps due to scheduling or instrumental constraints.

\subsection{Performance on Observational Data} \label{subsec:real-reconstruction}

To assess the ability of each model to generalize beyond synthetic data, we evaluate their performance on a real observational dataset of Sgr A*, as described in Section~\ref{subsec: observational data}. Importantly, none of the models were trained or fine-tuned on this dataset. There was no exposure to observational data during training, nor was any empirical prior incorporated. This evaluation therefore probes each model's capacity to transfer knowledge from a purely simulated training distribution to real-world astrophysical variability.

\begin{figure*}[ht]
\includegraphics[width=\textwidth]{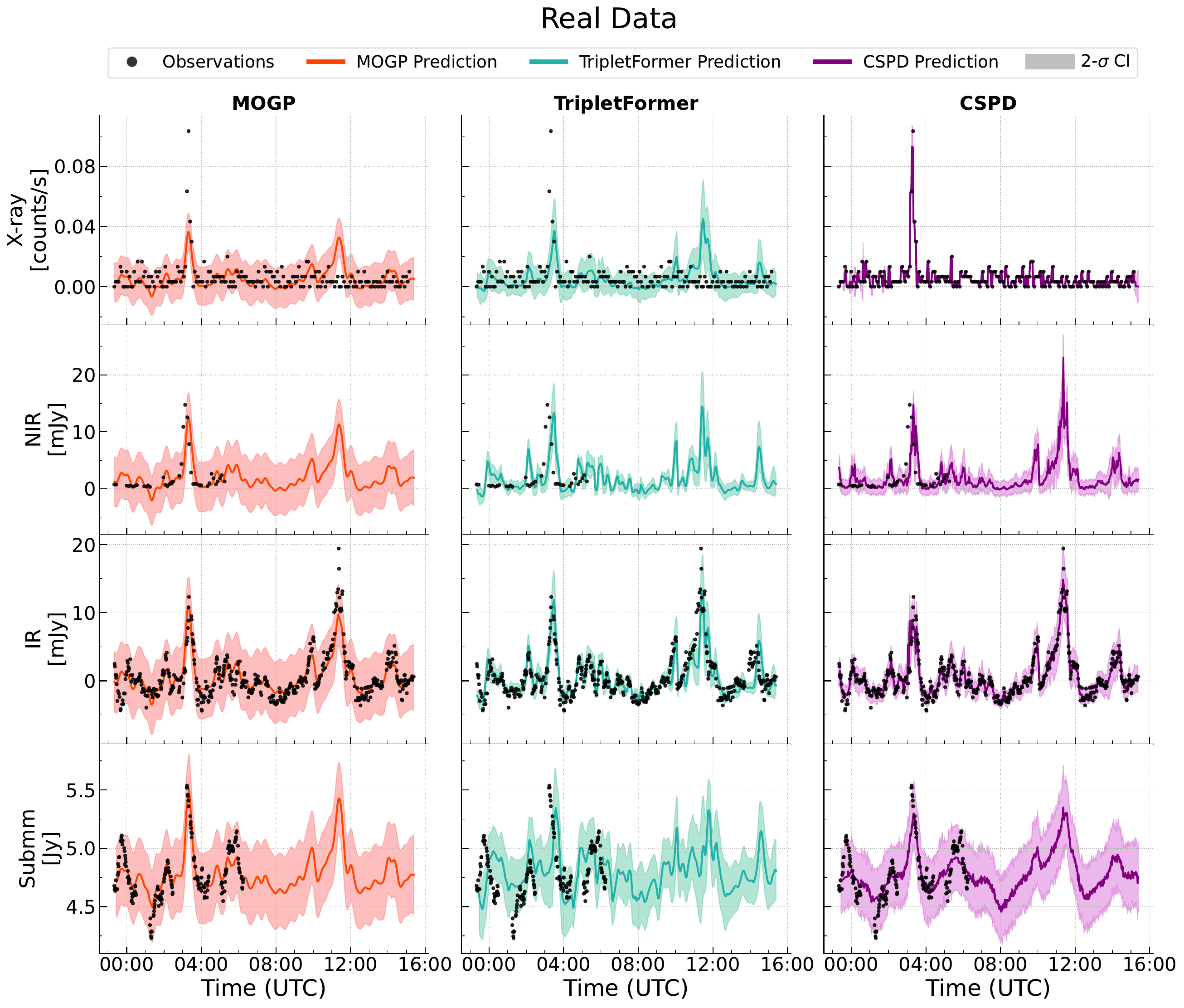}
\caption{Model interpolation results on real observational data of Sgr A* from Section~\ref{subsec: observational data}. Each column corresponds to a different model (MOGP, TripletFormer, CSPD), and each row represents a different wavelength band (X-ray, NIR, IR, submm). Observational data upon which the models are conditioned are shown as black circles. Solid lines denote the mean predictions of each model while shaded regions denote 2$\sigma$ credible intervals. While all models generally follow the observed data, the CSPD model produces sharper predictions with more selective uncertainty and avoids false-positive flares hallucinated by the other models.}
\label{fig:real data interpolation}
\end{figure*}

Each model is conditioned on all available observations and used to reconstruct the full multi-wavelength light curve. The results are shown in Figure~\ref{fig:real data interpolation}, where panels are arranged by model (left to right: MOGP, TripletFormer, CSPD) and by wavelength (top to bottom: X-ray, NIR, IR, submm).

Qualitatively, we find that the CSPD and TripletFormer models produce mean predictions that more closely follow the observational data compared to the MOGP baseline. This is especially evident in the NIR and IR bands, where both deep learning models (CSPD and TripletFormer) track rapid flares more effectively. In contrast, MOGP reconstructions appear smoother and less responsive to sharp features, a behavior attributable to our use of a RBF kernel, which tends to favor smooth interpolants. This is consistent with the model’s behavior on simulated data (see Section~\ref{subsec:simulated performances}).

A key distinction arises in the models’ predicted uncertainties. The MOGP model exhibits considerably broader credible intervals across all wavelengths, reflecting its more conservative Gaussian assumptions (e.g., \citealt{bonilla_multi-task_2007, rasmussen_gaussian_2008}). The CSPD model produces sharply defined uncertainty bounds, particularly in the X-ray band, where they may be overly narrow, potentially indicating overconfidence in regions with low signal-to-noise. The TripletFormer offers intermediate uncertainty estimates, narrower than the MOGP but generally wider than those of CSPD.

To quantitatively evaluate model performance on incomplete observations, we conducted two types of masking tests, following the procedure described in Section~\ref{subsec:missing data}. In the first, we randomly masked individual data points across all wavelengths and repeated this process multiple times to obtain averaged metrics, capturing typical reconstruction performance under unstructured missingness. In the second, we simulated burst missing data, where contiguous blocks of observations were removed to emulate periods of instrument downtime or observational gaps. These complementary tests allow us to assess model robustness under both sporadic and temporally clustered missing data, providing a more comprehensive picture of predictive reliability across the multi-wavelength time series. MSE and CRPS results across these tests are shown in Tables~\ref{tab:mse_realdata} and \ref{tab:crps_realdata}, respectively.

Across both the random and burst missing data tests, all models exhibit a consistent pattern in their quantitative performance. It is important to note that, unlike in the simulated data experiments, the real multi-wavelength observations are expressed in different physical units and cannot be normalized across bands. As a result, the sub-mm wavelength dominates the numerical scale of the MSE values, and the results should therefore be interpreted on a per-wavelength basis rather than through global averages. Notably, the sub-mm band remains particularly challenging, with all approaches showing higher errors compared to the other wavelengths, a trend consistent with that observed in the simulated data. Among the models, CSPD and MOGP achieve good overall performance, yielding slightly lower MSE and CRPS values across nearly all wavelengths and for both missingness patterns compared to TripletFormer. However, the margin of improvement remains modest, suggesting that while CSPD can offer a more consistent reconstruction quality, the problem remains intrinsically difficult across methods. Interestingly, the difference between random and burst missing data scenarios is relatively small, indicating that temporal clustering of missing observations does not drastically degrade model performance. This suggests that all three models possess a degree of temporal robustness in their predictions, maintaining comparable reconstruction accuracy even under structured data gaps.

Of particular interest is a region between 10:00 and 13:00 UTC, where both the MOGP and TripletFormer models appear to hallucinate a flare in the X-ray band at approximately 11:30 UTC, despite observed data indicating no such event. In contrast, the CSPD model maintains a prediction more closely aligned with the ground truth during this interval. While this qualitative observation suggests that the CSPD model may better balance inter-band correlations with wavelength-specific dynamics, we emphasize that these results are exploratory. Both the MOGP and transformer architectures rely heavily on modeling joint correlations across bands and may be overly influenced by strong flares in other bands (e.g., \citealt{goovaerts_geostatistics_1997, bonilla_multi-task_2007, alvarez_computationally_2011, yalavarthi_tripletformer_2023}). A rigorous assessment of false positives on real data would require controlled masking experiments or additional datasets, which are not currently available. Such testing is essential for quantitatively evaluating model reliability, particularly in cases such as Sgr A* where spurious flares could lead to misleading interpretations. We therefore leave a systematic investigation of false-positive behavior to future work, while noting that these preliminary observations highlight the importance of robust evaluation protocols that account for both temporal dynamics and cross-band correlations.

\begin{deluxetable}{cccc}[ht]
\tablecaption{MSE Across Real Data: Random and Burst Missing Tests} \label{tab:mse_realdata}
\tablewidth{0pt}
\tablehead{
\colhead{Wavelength} & \colhead{MOGP} & \colhead{TripletFormer} & \colhead{Diffusion}
}
\startdata
\multicolumn{4}{c}{\textbf{Random Missing Data}} \\
\hline
X-ray   & $1.007 \pm 0.042$ & $1.163 \pm 0.082$ & $1.064 \pm 0.151$ \\
NIR     & $1.042 \pm 0.039$ & $1.301 \pm 0.081$ & $1.146 \pm 0.041$ \\
IR      & $0.150 \pm 0.022$ & $0.814 \pm 0.125$ & $0.038 \pm 0.005$ \\
Sub-mm  & $394.139 \pm 0.363$ & $398.489 \pm 0.468$ & $393.534 \pm 0.265$ \\
\hline
\multicolumn{4}{c}{\textbf{Burst Missing Data}} \\
\hline
X-ray   & $1.081 \pm 0.220$ & $1.371 \pm 0.307$ & $1.118 \pm 0.283$ \\
NIR     & $1.095 \pm 0.213$ & $1.383 \pm 0.363$ & $1.082 \pm 0.199$ \\
IR      & $0.793 \pm 0.640$ & $1.896 \pm 0.472$ & $0.814 \pm 0.726$ \\
Sub-mm  & $395.179 \pm 3.109$ & $393.124 \pm 3.769$ & $392.509 \pm 1.776$ \\
\enddata
\end{deluxetable}


This behavior also highlights an important physical constraint of the underlying emission mechanisms: not all NIR flares have corresponding X-ray counterparts (e.g., \citealt{yusef-zadeh_alma_2017}). While increases in particle acceleration may produce enhanced synchrotron radiation in the NIR, the production of X-ray photons, via synchrotron self-Compton processes, is nonlinear and sensitive to the population of high-energy electrons near the synchrotron cutoff (e.g., \citealt{dodds-eden_two_2011, ponti_powerful_2017, do_unprecedented_2019, witzel_rapid_2021}). As such, even when the NIR data exhibits a clear flare, the corresponding X-ray signal may remain below the detection threshold due to photon noise and the intrinsic stochasticity of the synchrotron self-Compton process (e.g., \citealt{meyer_formal_2014}). The ability of the CSPD model to avoid hallucinating such spurious X-ray flares suggests that it has successfully internalized these physical relationships from the training data, learning a more complex generative structure.

Another interesting feature appears in the ~3:30 UTC flare, where the IR peak is well captured across all three models, but the predicted peaks in the X-ray, submm, and NIR bands exhibit a slight time delay relative to the observed data. This behavior may arise from several factors. First, it could reflect model limitations: none of the models explicitly encode inter-band time lags, and their smoothing assumptions, especially in sparsely sampled regions, may bias them toward temporally offset reconstructions. Second, our models were trained exclusively on simulated light curves derived from a single physical model, which may constrain their ability to generalize to more complex or varied inter-band timing relationships. Future efforts could benefit from incorporating models with physically motivated time-lag structures to capture such effects more explicitly.

Despite these differences, all three models exhibit diminished performance in the submm band, consistent with results observed on the simulated dataset. 

Overall, these results suggest that CSPD offers a promising approach for generalizing from Sgr A* light curve simulations to real observations. It maintains fidelity in both its predictive mean and uncertainty estimates across most modalities, while avoiding key failure modes observed in more rigid or correlation-driven architectures.

\subsection{Uncertainty Calibration} \label{subsec:Uncertainty Calibration}

\begin{figure}[ht]
\plotone{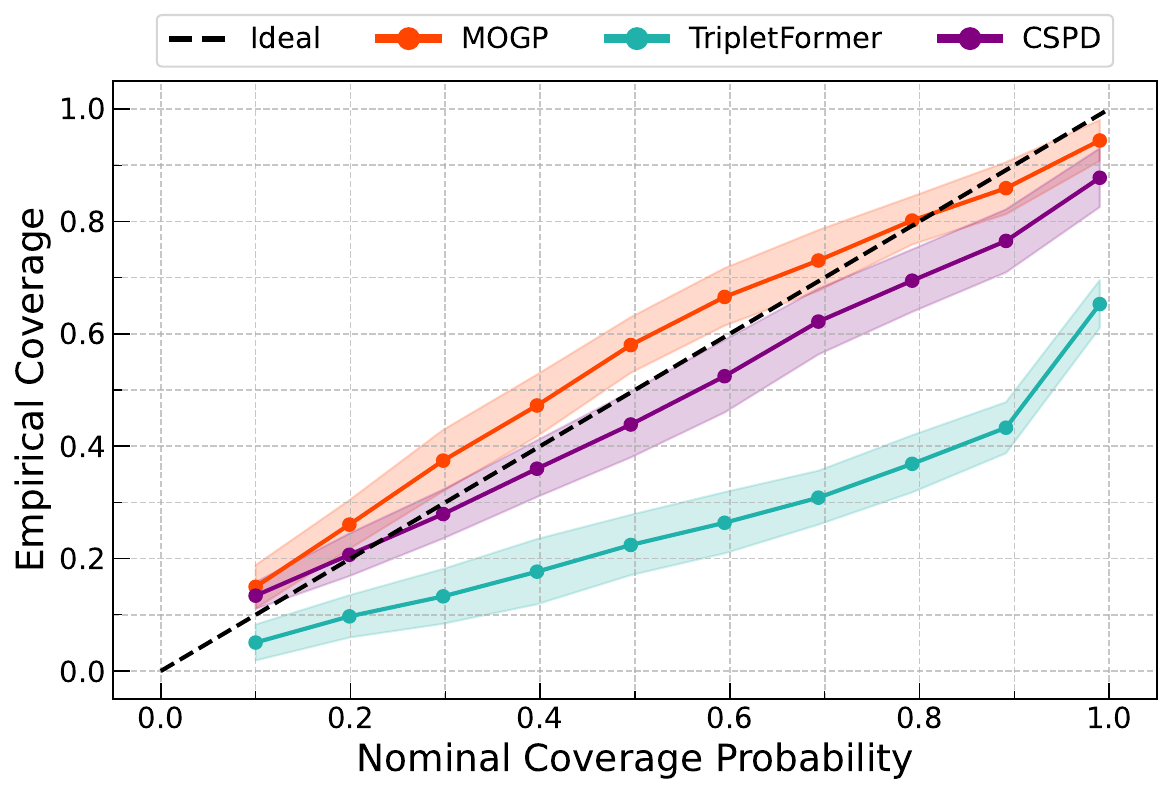}
\epsscale{1}
\caption{
Coverage analysis of predictive uncertainties across a range of nominal confidence levels, computed on real observational data. Each curve shows the empirical coverage achieved by a model as a function of the nominal coverage level, aggregated across all wavelength bands. The dashed diagonal line indicates perfect calibration. The MOGP (orange) shows the best overall calibration, closely following the ideal line. CSPD (purple) slightly underestimates uncertainty but remains well-calibrated, meanwhile the TripletFormer (teal) consistently undercovers.}
\label{fig:coverage_line}
\end{figure}

To evaluate how well each model's uncertainty estimates are calibrated across different confidence levels, we conduct a coverage analysis using the real observational dataset described in Section~\ref{subsec: observational data}. We assess empirical coverage across a range of nominal coverage probabilities, from 10\% to 99\%. For each trial, we randomly mask a subset of the observed flux values and ask each model to reconstruct the missing data along with its predictive uncertainty. We then compute the fraction of true values that fall within the model’s predicted credible intervals at each nominal level (e.g., \citealt{kuleshov_accurate_2018}). This process is repeated with different random masks to obtain stable estimates, and the final coverage curves are averaged across all four wavelength channels (e.g., \citealt{lakshminarayanan_simple_2017}). The results, presented in Figure~\ref{fig:coverage_line}, provide a model-agnostic assessment of calibration fidelity, a perfectly calibrated model would trace the identity line, where nominal and empirical coverage coincide (e.g., \citealt{ovadia_can_2019}).

The MOGP consistently shows the best alignment with the ideal coverage line. It slightly overcovers for nominal levels below 80\%, and transitions to mild undercoverage at higher thresholds, suggesting that while the model may slightly underestimate tail uncertainty, meaning its confidence in rare or extreme values, its predictions remain broadly reliable (e.g., \citealt{aigrain_gaussian_2023}). This performance reflects the well-understood behavior of GPs: their smooth priors yield robust, conservative uncertainty estimates, especially when data is sparse or irregular (e.g., \citealt{foreman-mackey_fast_2017}). The CSPD model also performs well, tracking just below the ideal line across most of the coverage range. Its credible intervals are slightly narrower than optimal, indicating a mild underestimation of uncertainty, but the deviation remains modest (e.g., \citealt{kuleshov_accurate_2018}). This result is encouraging given the model’s flexibility and its ability to learn complex, non-Gaussian noise structures from data, highlighting the promise of diffusion-based approaches for adaptive uncertainty quantification (e.g., \citealt{tashiro_csdi_2021, bilos_modeling_2023, mimikos-stamatopoulos_score-based_2024}).

By contrast, the TripletFormer model exhibits significant undercoverage across the full range of nominal levels. Its coverage curve remains well below the identity line, revealing a tendency toward overconfident predictions (e.g., \citealt{kuleshov_accurate_2018}). This miscalibration is likely rooted in the model’s simplified, parametric treatment of uncertainty and the limitations of its transformer-based architecture, which may struggle to capture rare, high-variance events (e.g., \citealt{yalavarthi_tripletformer_2023}). These results suggest that, while transformer models may excel in deterministic reconstruction tasks, as showcased by \cite{boersma_transformer_2024}, their native uncertainty estimates require more sophisticated probabilistic enhancements to achieve competitive calibration performance.

To complement this global evaluation, Appendix~\ref{fig:coverage_histogram} presents a band-wise breakdown of empirical coverage at a fixed 95\% ($2\sigma$) nominal level. This per-wavelength histogram analysis reveals that the MOGP achieves strong calibration across most bands but overestimates uncertainty in the IR, where its intervals become overly conservative (e.g., \citealt{rasmussen_gaussian_2008}). The CSPD model achieves consistent coverage across bands, with particularly good performance in the IR and submm. Meanwhile, all three models perform poorly in the NIR band, with significant undercoverage indicative of the extreme variability and flaring present in that channel (e.g., \citealt{dodds-eden_two_2011, fazio_multiwavelength_2018, do_unprecedented_2019, boyce_simultaneous_2019, boyce_multiwavelength_2022}). Taken together, these results highlight the importance of evaluating uncertainty calibration both globally and individually for each wavelength, and point to the need for models that can adapt their confidence estimates to the distinct statistical properties of each band (e.g., \citealt{ovadia_can_2019}).

To further evaluate the calibration of the models’ predictive uncertainties, we constructed Probability Integral Transform (PIT) histograms for each wavelength channel. The PIT histogram assesses how well the predicted distributions capture the variability of the true observations: ideally, a uniform distribution indicates perfectly calibrated uncertainties, while deviations from uniformity reveal under- or over-confidence (e.g., \citealt{david_probability_1948}). Figure~\ref{fig:pit_ir} shows an example PIT histogram specifically for the IR band, although other wavelengths exhibit similar behaviour. All three models exhibit pronounced peaks near 0 and 1, indicating that their predictive distributions frequently fail to encompass extreme observed values. Notably, the TripletFormer model displays the largest peaks, followed by CSPD, and then MOGP, which shows the smallest peaks. This ordering is consistent with our previous observations on coverage and uncertainty calibration: MOGP provides the most reliable uncertainty estimates, CSPD is moderately calibrated, and the TripletFormer tends to be overconfident. These results reinforce our earlier conclusions regarding the relative fidelity of the models’ uncertainty predictions.

While the coverage analysis provides valuable insights into the overall calibration of each model, it remains a relatively coarse, aggregate measure. A more thorough assessment would include additional diagnostics such as interval score evaluations, and out-of-distribution tests to systematically quantify over- or under-confidence across different flux levels and temporal regimes (e.g., \citealt{kuleshov_accurate_2018, kuleshov_calibrated_2022}). Given the limited size and sparsity of the available observational data, conducting such detailed analyses is currently challenging. Therefore, we consider these complementary evaluations an important direction for future work, which will help establish a more complete understanding of each model’s uncertainty behavior, particularly in extreme events or poorly sampled wavelengths.

\subsection{Limitations}
\label{subsec:limitations}
Our study is subject to several limitations that may influence the model comparisons and overall findings. The MOGP and TripletFormer baselines were intentionally implemented in a minimalist configuration. For the MOGP, we used a fixed RBF kernel without exploring more sophisticated alternatives, such as Matérn or spectral mixture kernels, which might better capture the complex astrophysical variability (e.g., \citealt{parra_spectral_2017}, \citealt{aigrain_gaussian_2023}, \citealt{paugnat_new_2024}). Furthermore, while we conducted hyperparameter tuning for the Tripleformer, our exploration did not include a variety of temporal encodings, which could have a significant impact on performance with irregularly sampled data (e.g., \citealt{kim_continuous-time_2024}). Consequently, the performance gains reported for our method should be interpreted in the context of these specific baseline implementations. A more comprehensive comparison with highly tuned baselines would be a valuable and warranted direction for future work.

Another limitation of our CSPD formulation is that the forward-process kernel 
\(k(t_i, t_j)\) encodes only temporal correlations within each band. We did not 
implement a structured multi-output kernel that could explicitly encode 
wavelength-dependent correlations (e.g., intrinsic coregionalization models). 
Instead, cross-band dependencies are learned implicitly by the neural network. 
While this design choice improves computational tractability, it reduces the 
degree of physical interpretability at the prior level, since the CSPD prior 
assumes independence across bands. As a result, the model may attribute 
astrophysical correlations between bands entirely to the learned denoiser rather 
than to a physically motivated prior. Future work could extend the CSPD 
framework with structured multi-output kernels, thereby embedding prior 
knowledge of expected cross-band variability directly into the generative model.

Relatedly, we did not perform explicit diagnostics linking the learned behavior 
to physical timescales, such as X-ray flare durations or submm lag structures. 
Such tests would help verify that the model captures astrophysically meaningful 
dependencies, beyond reproducing statistical correlations. Without these 
diagnostics, there remains some uncertainty as to whether the reconstructed 
cross-band variability reflects true physical processes or artifacts of the 
neural network’s learned correlations.

Additional limitations concern the simulated data itself. All model training, validation, and selection are based on a single simulation family from \cite{witzel_rapid_2021}, which encodes specific assumptions about cross-band couplings, such as treating the X-ray as a low-pass filtered version of the NIR. While the large ensemble of 16,350 sequences and cadence-matched masking improve robustness, the simulations do not cover the full diversity of plausible Sgr~A* variability. Consequently, the models may learn simulator-specific inductive biases and exploit particular temporal or spectral correlations, potentially limiting generalization to real data or alternative physical scenarios. We deliberately focused on a single, well-characterized simulator to clearly showcase the application of diffusion models to astrophysical time series interpolation. Addressing broader generalization, including “leave-physics-out” or cross-simulator stress tests, is an important avenue for future work to quantify the models’ susceptibility to simulator-specific biases.

\section{Conclusion} \label{sec:conclusion}

This work contributes directly to the scientific interpretation of Sgr A*'s variability. Observational campaigns of the Galactic Center are often sparse, irregularly sampled, and heterogeneous across wavelengths, making it difficult to constrain inter-band correlations and temporal structures that underpin physical models. By enabling reliable interpolation with quantified uncertainties, our approach helps extract more information from incomplete datasets and supports the development of data-driven priors for future analysis. Ultimately, this framework advances our ability to study the multi-scale variability of accreting BHs, even in the absence of more complete or higher-cadence observations.

While the results presented here demonstrate that diffusion-based models like CSPD can generalize reasonably well from simulated to real light curve data, there remains a clear gap in performance, particularly in more challenging bands such as submm. Improving generalization to real data is a key next step in the development of robust Sgr A* and other astronomical time series reconstruction models. 

The ability to reliably interpolate and quantify uncertainties in multi-wavelength light curves opens several avenues for advancing our understanding of Sgr A*. For instance, our proposed methods could be directly applied to constrain key physical parameters, such as the spectral index, by providing more complete simultaneous multi-wavelength coverage. Quantifying time delays between different emission bands, crucial for probing the size and dynamics of the emission region, would also benefit significantly from denser and more robustly interpolated light curves. Furthermore, enhanced data density could lead to more accurate measurements of the source's power spectrum, revealing characteristic timescales of variability. Beyond Sgr A*, this framework has broad applicability across time series astronomy, including the study of active galactic nuclei flare mechanisms, tidal disruption events, and even exoplanet transits, where incomplete or sparsely sampled observations often limit interpretation.

A compelling direction for future work is the inclusion of more diverse training data. Given that the models in this work were primarily trained on data drawn from a single simulation dataset, there is a recognized potential for the learned interpolation behavior, and thus subsequent conclusions, to be subtly influenced or even skewed towards the intrinsic characteristics of that training distribution (e.g., \citealt{huppenkothen_constructing_2023}). Therefore, combining phenomenologically generated simulations with physical general relativistic magnetohydrodynamic-based light curves, which have been shown to better reproduce the variability and morphology of accreting BHs, becomes even more critical for robust generalization (e.g., \citealt{chan_fast_2015, ball_particle_2016, chatterjee_general_2021}). Incorporating historical observational datasets directly into training, when available, can also help expose the model to realistic sampling irregularities and complex noise characteristics beyond simple homoscedasticity, such as heteroscedasticity or correlated noise, bridging the simulation-to-observation domain gap (e.g., \citealt{fortuin_gp-vae_2020}). Additionally, recent work by \cite{fagin_latent_2024} demonstrates the potential of on-the-fly simulation during training, in which new synthetic light curves are continuously generated during training, enabling models to encounter a wider diversity of physically plausible behaviors.

An alternative strategy is to maintain a simulation-only training set but guide inference with a learned observational prior. This can be done by training a diffusion model to encode the structure of real Sgr A* light curves, similar to the approach of \cite{adam_posterior_2022}, who introduced score-based priors for source galaxy reconstruction in gravitational lensing. In their work, a diffusion model is trained on high-resolution galaxy images to learn a flexible prior distribution. This prior is then combined with a likelihood term using a reverse-time stochastic differential equation solver to draw samples from the posterior. A similar approach in the time-domain context could help ensure that predictions remain consistent with real behavior of Sgr A* even when trained only on simulations.

Finally, recent advances in domain adaptation, self-supervised learning, and foundational models offer promising directions for improving model generalization from simulations to real observational data. Techniques such as adversarial feature alignment (e.g., \citealt{ganin_domain-adversarial_2016}), representation matching (e.g., \citealt{tzeng_adversarial_2017}), and contrastive learning (e.g., \citealt{chen_simple_2020}) have been successfully applied to reduce distributional shifts between synthetic and real-world domains. In the time series domain, \cite{yoon_time-series_2019} introduced TimeGAN, which combines adversarial training with recurrent architectures to synthesize realistic sequences, while \cite{german-morales_transfer_2025} demonstrate that foundational models, when adapted with efficient low-rank fine-tuning techniques, can achieve strong generalization in time series forecasting tasks, even in zero-shot settings where test distributions differ significantly from training data.

Altogether, these approaches reflect a growing need to move beyond purely synthetic modeling pipelines and toward hybrid frameworks that incorporate empirical structure, domain-specific priors, and physical realism in an uncertainty-aware framework.

\section*{Acknowledgments}

I thank Gunther Witzel for generously sharing the simulated light curves from his work and for insightful discussions about their use. I am grateful to Hadrien Paugnat for introducing me to MOGPs and providing guidance that helped shape the direction of this work. I also thank Joshua Fagin for stimulating conversations on time series modeling in astrophysics and the role of GPs in that context. I am indebted to Alex Adam for valuable discussions on diffusion models and score-based inference, which were instrumental in developing this approach. Finally, I thank Koushik Chatterjee for engaging discussions on GRMHD simulations and their relevance to modeling Sgr A* variability.

\section*{Data Availability} \label{sec:cite}

All methods and data used in this paper are available at \href{https://github.com/GabrielSasseville01/SgrA_Interpolation}{\faGithubSquare  GabrielSasseville01} and at the \dataset[Zenodo]{https://doi.org/10.5281/zenodo.17489387} repository.

\appendix
\restartappendixnumbering

\section{Hyperparameter Tuning}
\label{appendix:hyperparameter}

This appendix outlines the hyperparameter tuning strategy employed for the TripletFormer and CSPD models. For each model, we conduct a systematic grid search over key architectural parameters. Each configuration is independently trained for a maximum of 12 hours using one NVIDIA V100 GPU with 16GB of memory. Each job is run with access to 32GB of RAM. The tables \ref{tab:hyperparameters triplet} and \ref{tab:hyperparameters cspd} summarize the validation losses associated with each hyperparameter combination, for the Tripletformer and CSPD, respectively, with the best-performing configurations highlighted in bold.

\begin{deluxetable}{cccc c}[ht]
\tablecaption{TripletFormer Hyperparameter Combinations with their Corresponding Validation Loss \label{tab:hyperparameters triplet}}
\tablewidth{0pt}
\tablehead{
\colhead{\textbf{IMAB}} & \colhead{\textbf{CAB}} & \colhead{\textbf{Decoder}} & \colhead{\textbf{Number of}} & \colhead{\textbf{Validation}} \\
\colhead{\textbf{Dimension}} & \colhead{\textbf{Dimension}} & \colhead{\textbf{Dimension}} & \colhead{\textbf{Reference Points}} & \colhead{\textbf{Loss}}
}
\startdata
64  & 256  & 128  & 128 & 0.390699 \\
64  & 256  & 128  & 256 & 0.403597 \\
64  & 256  & 256 & 128 & 0.428478 \\
64  & 256  & 256 & 256 & 0.468442 \\
64  & 128  & 128  & 128 & 0.493111 \\
64  & 128  & 128  & 256 & 0.470652 \\
64  & 128  & 256 & 128 & 0.487339 \\
64  & 128  & 256 & 256 & 0.450617 \\
\textbf{128}  & \textbf{256}  & \textbf{128}  & \textbf{128} & \textbf{0.341686} \\
128  & 256  & 128  & 256 & 0.366206 \\
128  & 256  & 256 & 128 & 0.366558 \\
128  & 256  & 256 & 256 & 0.358992 \\
128  & 128  & 128  & 128 & 0.460786 \\
128  & 128  & 128  & 256 & 0.507229 \\
128  & 128  & 256 & 128 & 0.531182 \\
128  & 128  & 256 & 256 & 0.484461 \\
\enddata
\tablecomments{The configuration in bold corresponds to the configuration with best validation loss, thus used for subsequent experiments.}
\end{deluxetable}

\begin{deluxetable}{cccc c}[ht]
\tablecaption{CSPD Hyperparameter Tuning Combinations with their Corresponding Validation Loss \label{tab:hyperparameters cspd}}
\tablewidth{0pt}
\tablehead{
\colhead{\textbf{Layers}} & \colhead{\textbf{Channels}} & \colhead{\textbf{Number of}} & \colhead{\textbf{Diffusion Embedding}} & \colhead{\textbf{Validation}} \\
 &  & \colhead{\textbf{Heads}} & \colhead{\textbf{Dimension}} & \colhead{\textbf{Loss}}
}
\startdata
4  & 64  & 8  & 128 & 47.6492 \\
4  & 64  & 8  & 256 & 47.9842 \\
4  & 64  & 16 & 128 & 72.0357 \\
4  & 64  & 16 & 256 & 59.4039 \\
4  & 128  & 8  & 128 & 46.1666 \\
4  & 128  & 8  & 256 & 46.4126 \\
4  & 128  & 16 & 128 & 47.4152 \\
4  & 128  & 16 & 256 & 47.3068 \\
8  & 64  & 8  & 128 & 48.3733 \\
8  & 64  & 8  & 256 & 48.8815 \\
8  & 64  & 16 & 128 & 72.1815 \\
8  & 64  & 16 & 256 & 68.8929 \\
8  & 128  & 8  & 128 & 45.6741 \\
\textbf{8}  & \textbf{128}  & \textbf{8}  & \textbf{256} & \textbf{45.3468} \\
8  & 128  & 16 & 128 & 65.5916 \\
8  & 128  & 16 & 256 & 64.0322 \\
\enddata
\tablecomments{The configuration in bold corresponds to the configuration with best validation loss, thus used for subsequent experiments.}
\end{deluxetable}

\clearpage

\section{Wavelength-Specific Results}

\begin{figure*}[ht]
\includegraphics[width=\textwidth]{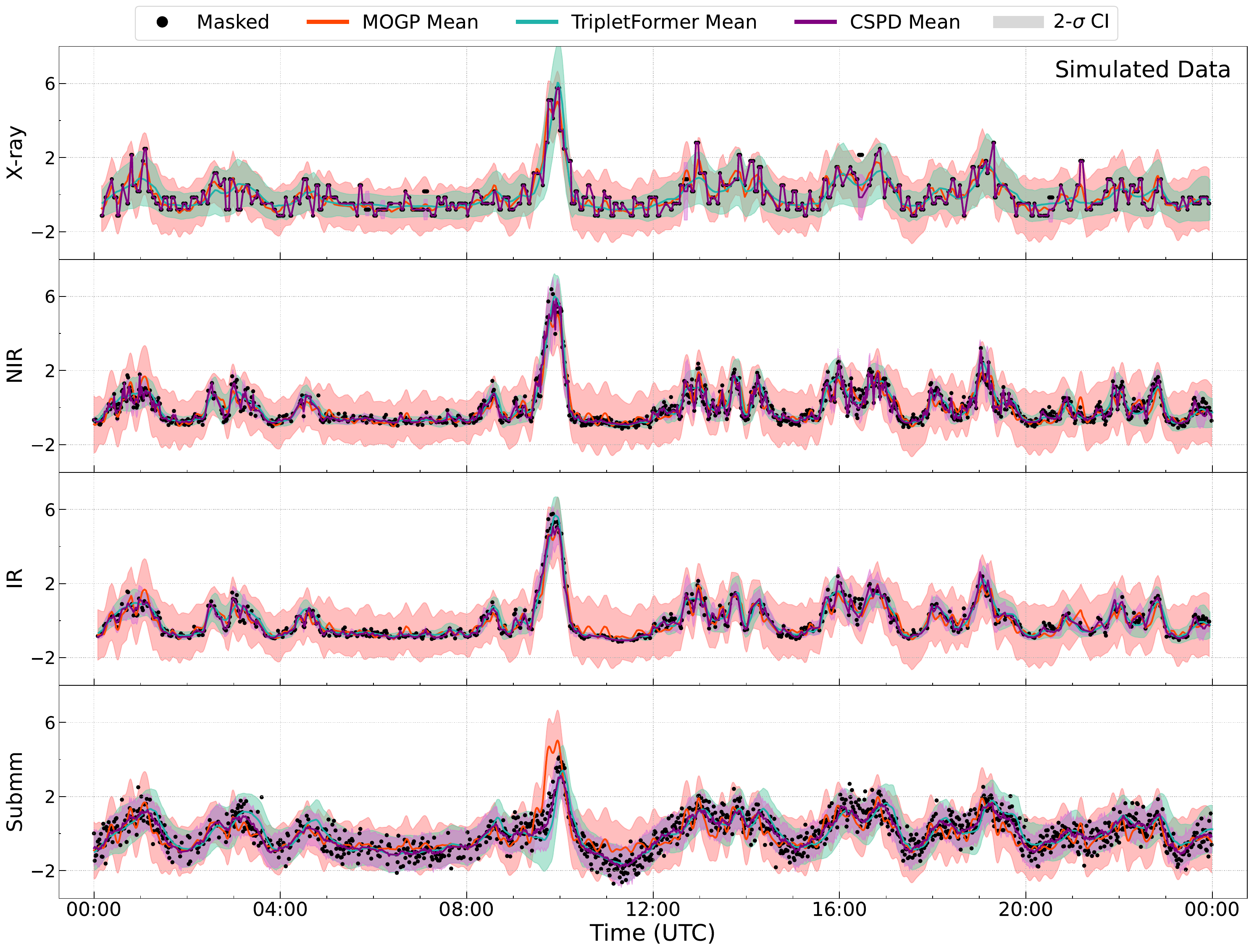}
\caption{Example interpolation of the simulated multivariate light curve in Figure~\ref{fig:simulationdata} by each model. Each row represents a different wavelength band: X-ray (2-8 keV), NIR (2.2 $\mu$m), IR (4.5 $\mu$m), submm (340 GHz). The ground truth flux values to be predicted are shown as black circles. Red, teal and purple lines represent the MOGP, TripletFormer and CSPD models' mean prediction. Shaded regions represent 2$\sigma$ credible intervals. All models successfully recover the underlying signal, but the CSPD model captures sharper variability and exhibits tighter uncertainty bounds compared to the smoother, more conservative predictions of the MOGP and TripletFormer.}
\label{fig:simulated data interpolation}
\end{figure*}

\begin{figure}[ht]
\plotone{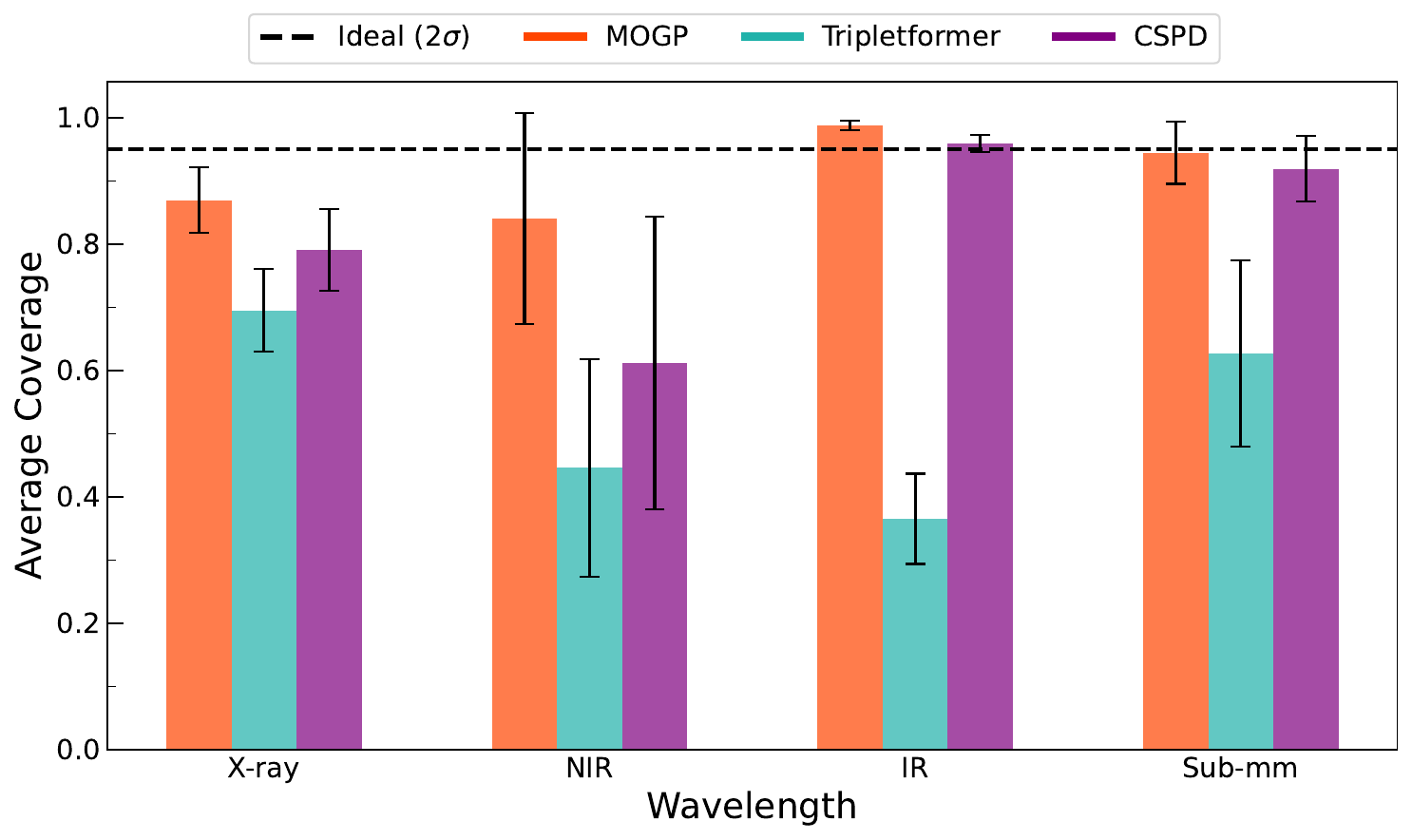}
\epsscale{1}
\caption{
Empirical coverage at the 95\% ($2\sigma$) confidence level, computed independently for each wavelength. Each bar represents the average coverage achieved by one model when tasked with reconstructing masked observations across multiple randomized trials. The dashed horizontal line indicates the ideal 95\% coverage level expected from a perfectly calibrated model. The MOGP and CSPD models achieve reasonably well-calibrated uncertainty estimates across most bands, while the TripletFormer consistently undercovers. All models struggle in the NIR band, highlighting the need for wavelength-specific evaluation.}

\label{fig:coverage_histogram}
\end{figure}

\begin{figure}[ht]
\plotone{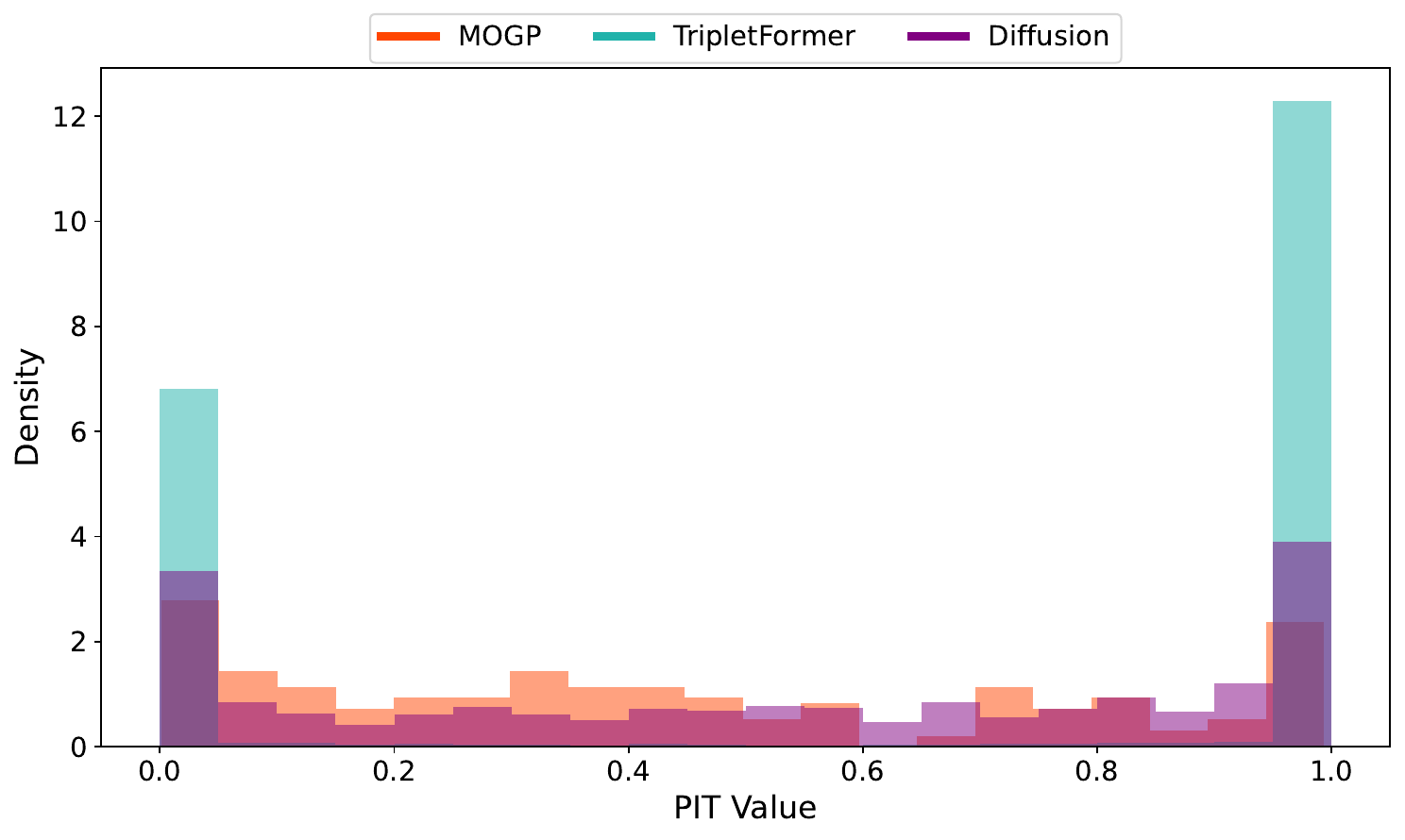}
\epsscale{1}
\caption{
PIT histogram for the IR band on real data. Peaks at 0 and 1 indicate deviations from uniformity, with TripletFormer showing the strongest peaks, followed by CSPD, and MOGP showing the least. This reflects the relative calibration of the models, consistent with our previous results: MOGP is best calibrated, followed by CSPD, then TripletFormer.}

\label{fig:pit_ir}
\end{figure}

This appendix presents additional figures showcasing the performance of each model individually across all four wavelengths. These include the full 24 hour version of the multi-wavelength interpolation Figure \ref{fig:simulatedinterpolation per model} in Figure~\ref{fig:simulated data interpolation}, wavelength-specific coverage histograms evaluating uncertainty calibration at the 95\% confidence level in Figure~\ref{fig:coverage_histogram}, as well as a PIT histogram for the IR band on real data in Figure~\ref{fig:pit_ir}. These plots provide a more detailed view of model behavior in each channel and highlight both strengths and limitations that may not be fully captured in the aggregate analyses. We also include further results for the missing data tests on real data in Table~\ref{tab:crps_realdata}.

\begin{deluxetable}{cccc}[ht]
\tablecaption{CRPS Across Real Data: Random and Burst Missing Tests} \label{tab:crps_realdata}
\tablewidth{0pt}
\tablehead{
\colhead{Wavelength} & \colhead{MOGP} & \colhead{TripletFormer} & \colhead{Diffusion}
}
\startdata
\multicolumn{4}{c}{\textbf{Random Missing Data}} \\
\hline
X-ray   & $0.508 \pm 0.010$ & $0.516 \pm 0.012$ & $0.864 \pm 0.045$ \\
NIR     & $0.524 \pm 0.008$ & $0.627 \pm 0.013$ & $0.843 \pm 0.016$ \\
IR      & $0.227 \pm 0.009$ & $0.536 \pm 0.040$ & $0.161 \pm 0.009$ \\
Sub-mm  & $19.042 \pm 0.014$ & $19.248 \pm 0.012$ & $0.985 \pm 0.001$ \\
\hline
\multicolumn{4}{c}{\textbf{Burst Missing Data}} \\
\hline
X-ray   & $0.524 \pm 0.057$ & $0.618 \pm 0.072$ & $0.884 \pm 0.112$ \\
NIR     & $0.536 \pm 0.054$ & $0.685 \pm 0.086$ & $0.816 \pm 0.078$ \\
IR      & $0.466 \pm 0.154$ & $0.771 \pm 0.211$ & $0.703 \pm 0.159$ \\
Sub-mm  & $19.077 \pm 0.074$ & $19.159 \pm 0.063$ & $0.984 \pm 0.003$ \\
\enddata
\end{deluxetable}

\newpage

\clearpage
\bibliography{SgrA}{}
\bibliographystyle{aasjournal}



\end{document}